\documentclass[aps,pra,showpacs,twocolumn,10pt,superscriptaddress]{revtex4-1}
\usepackage{graphicx}
\usepackage{amssymb}
\usepackage{amsmath}
\usepackage{bm}
\usepackage{hyperref}
\usepackage{color,soul}

\DeclareMathOperator\arctanh{arctanh}
\allowdisplaybreaks

\begin{document}

\title{Dynamical density and spin response of Fermi arcs\\
and their consequences for Weyl semimetals}

\author{Sayandip Ghosh}
\affiliation{Institute of Theoretical Physics, Technische Universit{\"a}t Dresden, 01062 Dresden, Germany}
\affiliation{Istituto Italiano di Tecnologia, Graphene Labs, Via Morego 30, 16163 Genova, Italy}

\author{Carsten Timm}
\email{carsten.timm@tu-dresden.de}
\affiliation{Institute of Theoretical Physics, Technische Universit{\"a}t Dresden, 01062 Dresden, Germany}

\date{March 13, 2020}

\begin{abstract}
Weyl semimetals exhibit exotic Fermi-arc surface states, which strongly affect their electromagnetic properties. We derive analytical expressions for all components of the composite density-spin response tensor for the surfaces states of a Weyl-semimetal model obtained by closing the band gap in a topological insulating state and introducing a time-reversal-symmetry-breaking term. Based on the results, we discuss the electromagnetic susceptibilities, the current response, and other physical effects arising from the density-spin response. We find a magnetoelectric effect caused solely by the Fermi arcs. We also discuss the effect of electron-electron interactions within the random phase approximation and investigate the dispersion of surface plasmons formed by Fermi-arc states. Our work is useful for understanding the electromagnetic and optical properties of the Fermi arcs.
\end{abstract}

\maketitle

\section{Introduction}

Following the discovery of several candidate materials \cite{Xu2015,Lv2015,Yang2015,Huang2015,Nakatsuji2015,Xiong2015,Liu2016,Xu2016,Hirschberger2016, Nayak2016,Ikhlas2017,Li2017,Liu2018,Wang2018,Hutt2018}, the study of Weyl semimetals (WSMs) has been an area of intensive research activities in recent years, see Refs.\ \cite{Hasan2017,Burkov2018,Armitage2018} for recent reviews. The band structure of WSMs is characterized by linear band-crossing points, so-called Weyl nodes, which act as sources or sinks of Berry curvature. The corresponding Berry charge, the chirality, is quantized to an integer value. The Weyl nodes always appear in pairs of opposite chirality and are separated in momentum (energy) when time-reversal (inversion) symmetry is broken. The nontrivial topology of the band structure and the quasirelativistic nature of the quasiparticles (the ``Weyl fermions'') gives rise to a plethora of exotic transport properties such as the magnetoelectric effect \cite{Zyuzin2012,Martin-Ruiz2019}, which signifies a electric-polarization response to magnetic fields and a magnetization response to electric fields, the anomalous Hall effect \cite{Goswami2013}, the dynamical chiral magnetic effect \cite{Ma2015}, and negative magnetoresistance \cite{Reis2016}. Transport in WSMs has recently been reviewed in Refs.~\cite{Wang2017,Gorbar2018,Hu2019}.

Furthermore, WSMs host unusual surface states known as Fermi arcs (FAs). These are disjointed segements of two-dimensional Fermi contours that connect the projections of a pair of bulk Weyl nodes of opposite chiralities into the surface Brillouin zone (BZ). They can be observed by angle-resolved photoemision spectroscopy \cite{Xu2015,Lv2015,Yang2015,Liu2016,Xu2016} and quasiparticle interference \cite{Zheng2016,Batabyal2016,Inoue2016,Gyenis2016}. The FAs are the most stringent signature of WSMs and thus were first used as smoking-gun evidence for the existence of WSMs. These states are topologically protected against weak disorder, show spin polarization and spin-momentum locking \cite{Xu2014,Xu2016a,Sakano2017}, and exhibit transport properties markedly different from the bulk~\cite{Potter2014,Moll2016}. 

The electromagnetic and transport properties exhibited by WSMs has been studied theoretically using semiclassical transport theory \cite{Son2013,Ma2015,Zhong2016,Breitkreiz2019}, field-theoretic approaches \cite{Goswami2013,Huang2017}, and the Kubo formalism \cite{Chang2015,Tabert2016,Roy2016}. The dynamical density, spin, and current responses to inhomogenous and time-dependent external fields have been investigated for Weyl fermions in the bulk, using Kohn-Luttinger-type ($\mathbf{k}\cdot\mathbf{p}$) models \cite{Thakur2018,Zhou2018,Ghosh2019}. The dynamical current response provides information about transport properties such as the chiral magnetic effect and the optical conductivity \cite{Zhou2018}. On the other hand, exploring the coupled density and spin response reveals the existence of novel collective excitations, i.e, spin plasmons, which provide another experimental signature of WSMs \cite{Ghosh2019}. While the surface plasmon excitations of the FA states have been investigated \cite{Song2017,Hofmann2016,Andolina2018,Losic2018,Gorbar2019}, an unified study of the density, spin, and current response of the FA states has been lacking but is desirable as it reflects the rich physics of the FA states.

In this paper, we investigate the response of the FAs to inhomogeneous and time-dependent electric and magnetic perturbations by analyzing all components of the composite density-spin linear-response tensor. We calculate the evanescent wave functions for the FAs and obtain analytical expressions for all components of the density-spin response tensor. We explore its observable consequences, which reveals the existence of a chiral magnetoelectric effect due to the FAs. For comparison and completeness, we also discuss the corresponding quantities for WSM bulk states. Based on the response tensor, we investigate the impact of the electron-electron interaction on the response functions for FA as well as bulk states within the random phase approximation (RPA). We also examine the spectrum of surface density excitations of FAs: the ``Fermi-arc plasmons.''

The remainder of this paper is organized as follows: in Sec.\ \ref{sec2}, we present our model and obtain the FA wave functions. Then, we introduce and calculate the dynamical response tensor in Sec.\ \ref{sec3} and investigate its manifestations in Sec.\ \ref{sec4}. The effect of the electron-electron interaction on the response function and the FA plasmon are discussed in Sec.\ \ref{sec5}. Finally, we summarize our results and draw conclusions in Sec.\ \ref{sec6}. Technical details and calculations are discussed in the appendices.

\section{Model}
\label{sec2}

We start with a four-band model used to describe three-dimensional topological insulators (TIs) of the Bi$_2$Si$_3$ family \cite{Zhang2009,Liu2010}. The low-energy effective Hamiltonian, regularized on a simple cubic lattice, is given by
\begin{align}
{\mathcal H}_0 &= \epsilon \sum_{\bf i} \Psi^{\dagger}_{\bf i}\, \sigma_0 \otimes \tau_{z}\, \Psi_{\bf i}
  - t \sum_{\langle{\bf i}, {\bf j}\rangle} \Psi^{\dagger}_{\bf i}\, \sigma_0 \otimes \tau_{z}\, \Psi_{\bf j}
  \nonumber \\
&{}+ i\lambda \sum_{\bf i} \Psi^{\dagger}_{\bf i} \left(\sigma_x \otimes \tau_{x}\, \Psi_{{\bf i} + \hat{\bf x}}
  + \sigma_y \otimes \tau_{x}\, \Psi_{{\bf i} + \hat{\bf y}}\right) \nonumber \\
&{}+ i\lambda_z \sum_{\bf i} \Psi^{\dagger}_{\bf i}\, \sigma_z \otimes \tau_{x}\, \Psi_{{\bf i} + \hat{\bf z}}
  + \text{H.c.} ,
\label{Hamil1}
\end{align}
where $\Psi_{\bf i}$ is a four-component fermion spinor operator, ${\bf i}$, ${\bf j}$ refer to lattice sites, ${\bf i} + {\bf l}$ denotes the nearest neighbor of site ${\bf i}$ in the ${\bf l}$ direction (${\bf l} = \hat{\bf x},\hat{\bf y},\hat{\bf z}$), $\epsilon$ and $t$ are the on-site energy and nearest-neighbor hopping amplitude, respectively, and $\lambda$ and $\lambda_z$ denote the spin-orbit coupling strengths in the $xy$-plane and along the $z$-direction, respectively. Here, $z$ is taken as the growth direction. $\sigma_l$ and $\tau_l$ are the $2 \times 2$ identity ($l=0$) and Pauli ($l=x,y,z$) matrices in spin and orbital space, respectively. The Hamiltonian $\mathcal{H}_0$ obeys time-reversal ($\mathcal{T}$) as well as inversion ($\mathcal{P}$) symmetry, whose representations are given by $\mathcal{T} = (i \sigma_y \otimes \tau_0)\, \mathcal{K}$ and $\mathcal{P} = \sigma_0 \otimes \tau_z$, respectively, with $\mathcal{K}$ denoting complex conjugation. Evidently, the two orbitals are of opposite parity.

The Hamiltonian describes a weak TI for $|\epsilon| < 2|t|$, a strong TI for $2|t| < |\epsilon| < 6|t|$, and a trivial insulator for $6|t| < |\epsilon|$ \cite{Khanna2014,Vazifeh2013}. In the following, we will consider the phase boundary between the topological and trivial insulator at $\epsilon = 6t$ with $t>0$, where the bulk gap closes at ${\bf k} = 0$ and Eq.\ (\ref{Hamil1}) describes a massless Dirac Hamiltonian.

A perturbation that breaks inversion or time-reversal symmetry results in the doubly degenerate Dirac node at ${\bf k} = 0$ being separated into a pair of Weyl nodes, i.e., in the emergence of a WSM phase. The Hamiltonian for the WSM is given by $\mathcal{H}_W = \mathcal{H}_0 + \mathcal{H}_P$, with
\begin{equation}
\mathcal{H}_P = \sum_{\bf i} \Psi^{\dagger}_{\bf i}
  \left[ b_0\, \sigma_0 \otimes \tau_{x}
  + {\bf b} \cdot (\sigma_x,\sigma_y,\sigma_z) \otimes \tau_{0} \right] \Psi_{\bf i} .
\end{equation}
Here, the $b_0$ term obeys time-reversal symmetry but breaks inversion symmetry, whereas the reverse holds for the ${\bf b}$ term~\cite{Vazifeh2013}. This rather simple model exhibits the relevant physics, while at the same time allowing for an analytical and hence transparent and rigorous treatment of the dynamical response functions and their observable consequences.

\subsection{Bulk Weyl semimetals}

The Hamiltonian for the infinite WSM system reads, in momentum space,
\begin{align}
\mathcal{H}({\bf k}) &= M_{\bf k} \, \sigma_0 \otimes \tau_{z}
  + 2\lambda \left(\sin k_x \, \sigma_x \otimes \tau_{x} + \sin k_y \, \sigma_y \otimes \tau_{x}
  \right) \nonumber \\
&{}+ 2\lambda_z \sin k_z \, \sigma_z \otimes \tau_{x} \nonumber \\
&{}+ b_0\, \sigma_0 \otimes \tau_{x} + {\bf b}\cdot (\sigma_x,\sigma_y,\sigma_z) \otimes \tau_{0} ,
\label{hamil2}
\end{align}
where $M_{\bf k} = \epsilon - 2t \sum_{\alpha} \cos k_{\alpha}$ is the ${\bf k}$-dependent mass parameter. We take $\epsilon = 6t$ so that when both $\mathcal{P}$ and $\mathcal{T}$ symmetries are present ($b_0 = 0, {\bf b} = 0$), the Hamiltonian yields a doubly degenerate Dirac node at ${\bf k} = 0$. When time-reversal symmetry is broken by ${\bf b}\neq 0$, the Dirac node splits into a pair of Weyl nodes separated in momentum, whereas breaking inversion symmetry by $b_0\neq 0$ results in Weyl nodes separated in energy. In the following, we consider the former scenario. Without loss of generality, we take ${\bf b} = b_y\, \hat{\bf y}$, which yields a pair of Weyl nodes at ${\bf k} = \pm (b_y/2\lambda)\, \hat{\bf y}$. The eigenenergies and eigenstates in the vicinity of the nodes are discussed in Appendix~\ref{app_bulk}.

\begin{figure}[tb]
\includegraphics[width=1.0\columnwidth]{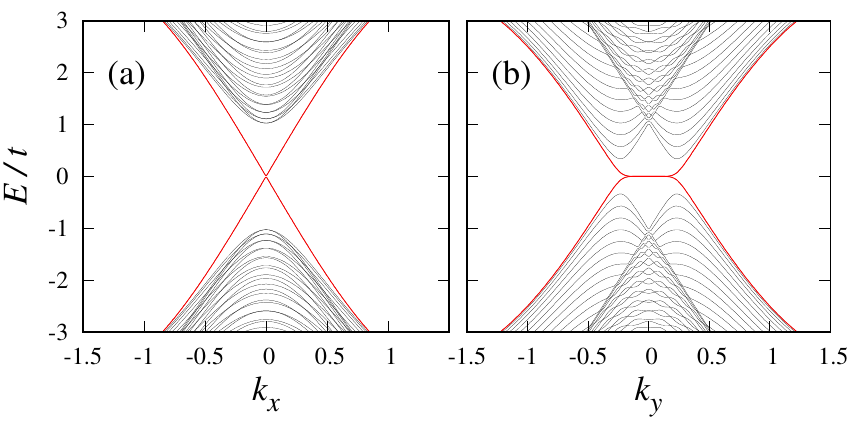}
\caption{Energy dispersion for a WSM on a slab of finite width in the $z$-direction, (a) for varying $k_x$ and $k_y=0$ and (b) for $k_x=0$ and varying $k_y$. The Weyl nodes are separated by $(b_y/\lambda)\, \hat{\bf y}$. The surface bands are shown in red. The black curves denote bulk-type bands, which are discrete due to the finite thickness ($50$ layers). The parameters are $b_y = t$ and $\lambda = \lambda_z = 2t$.}
\label{fig_dispersion} 
\end{figure}

\subsection{Fermi-arc states}

Our main interest is in the surface states of WSMs. In three-dimensional TIs, the bulk is gapped. The surface states are well separated from the bulk states and exist at each surface as mid-gap states. Each surface has a single two-dimensional Dirac cone which is called a ``helical metal'' owing to its spin-momentum locking \cite{Zhang2009,Wu2006,*Konig2008,*Qi2011,*Zhang2012}. In WSMs, even though the bulk is gapless, surface states can exist between the Weyl nodes in regions of the dispersion where there are no bulk states. For inversion-symmetric WSMs as considered here, states exist at surfaces that are not orthogonal to the momentum vector $(b_y/\lambda)\, \hat{\bf y}$ separating the nodes. Here, we consider the $(001)$ surface, perpendicular to the growth direction. The energy dispersion is shown in Fig.\ \ref{fig_dispersion}. Evidently, at low energies, the surface states form a band that disperses linearly in $k_x$, is flat along $k_y$, and only exists between the Weyl nodes. These are the FA states.

The essential idea for the calculation of the FA wave functions is to solve the Schr\"odinger equation for the Hamiltonian in Eq.~(\ref{hamil2}), replacing $k_z \rightarrow - i\, \partial/\partial z$ in order to treat the $z$-direction in real space. The wave function is separated into factors that are evanescent in $z$ and periodic in $(x,y)$. The proper boundary condition is obtained by modeling the vacuum as a large-gap insulator. Details of the derivation are presented in Appendix~\ref{app_FA}.

The FA states have the chiral dispersions $E^{T}({\bf k}) = - 2\lambda k_x$ and $E^{B}({\bf k}) = 2\lambda k_x$ for the top and bottom surfaces, respectively. Their density of states is $\mathcal{N}^{T,B} = b_y/8\pi^2 \lambda^2$, which does not depend on energy due to the linear dispersion \footnote{Strictly speaking, the Weyl nodes are located at $\pm \arcsin(b_y/2\lambda)\, \hat{\bf y}$ and the dispersion of the FAs is $E^{T,B}({\bf k}) = \pm 2\lambda \sin k_x$. However, as evident from Fig.\ \ref{fig_dispersion}, linearity of the FA energy in $k_x$ is a good approximation at low energies.}. As noted above, they only exist for $-b_y/2\lambda \leq k_y \leq b_y/2\lambda$, i.e., between the projections of the Weyl nodes into the surface BZ. The wave functions for the FAs at the top and bottom surfaces read
\begin{widetext}
\begin{align}
\Phi^{T}({\bf k}_\parallel,z) &= \sqrt{\frac{b_y^2 - 4 \lambda^2 k_y^2}{4 b_y \lambda_z}}
  \left[ \frac{1 - i}{2} \left(\begin{array}{c}
    1 \\ -1 \\ 1 \\ -1
  \end{array}\right) e^{\frac{b_y - 2\lambda k_y}{2\lambda_z}z} + \frac{1 + i}{2} \left(\begin{array}{c}
    1 \\ 1 \\ -1 \\ -1
  \end{array}\right) e^{\frac{b_y + 2\lambda k_y}{2\lambda_z}z} \right]
    e^{i{\bf k}_\parallel \cdot {\bf r}_\parallel},
\label{wave_function_FA.1} \\
\Phi^{B}({\bf k}_\parallel,z) &= \sqrt{\frac{b_y^2 - 4 \lambda^2 k_y^2}{4 b_y \lambda_z}}
  \left[ \frac{1+i}{2} \left(\begin{array}{c}
    1 \\ -1 \\ -1 \\ 1
  \end{array}\right) e^{-\frac{b_y - 2\lambda k_y}{2\lambda_z}z} + \frac{1-i}{2} \left(\begin{array}{c}
    1 \\ 1 \\ 1 \\ 1
  \end{array}\right) e^{-\frac{b_y + 2\lambda k_y}{2\lambda_z}z} \right]
    e^{i{\bf k}_\parallel \cdot {\bf r}_\parallel} .
\label{wave_function_FA.2}
\end{align}
\end{widetext}  
Here, the basis is chosen as $\{ |{\uparrow}{\oplus}\rangle, |{\uparrow}{\ominus}\rangle, |{\downarrow}{\oplus}\rangle, |{\downarrow}{\ominus}\rangle \}$, where $\uparrow$, $\downarrow$ denotes the $z$-component of the spin and $\oplus$, $\ominus$ denotes the even- and odd-parity orbital, respectively. ${\bf k}_\parallel=(k_x,k_y)$ is the two-dimensional momentum in the surface BZ.

Evidently, the FA states are bound to the surfaces and decay into the bulk on a $k_y$-dependent length scale $2\lambda_z/(b_y \pm 2\lambda k_y)$. Thus, for $k_y$ near Weyl nodes, the FA states extend deep into the bulk \cite{Haldane2014,Michetti2017,Armitage2018,Andolina2018,Adinehvand2019}. The FA states are eigenstates of $\sigma_x \otimes \tau_x$ with eigenvalues $-1$ and $+1$ for the top and bottom surfaces, respectively. In other words, the effective FA Hamiltonian is $\mathcal{H}_{\rm FA}( {\bf k}_\parallel) = 2\lambda k_x\, \sigma_x \otimes \tau_x$. The qualitative form and evanescent structure of FA states does not depend on the specifics of the model and survives for more generic Hamiltonians as discussed in Appendix~\ref{app_generalization}.

\section{Dynamical response}
\label{sec3}

The response of a system to possibly inhomogeneous and time-dependent electromagnetic perturbations is described by a $4 \times 4$ density-spin linear-response tensor, each component of which is, in principle, a $2 \times 2 \times 2 \times 2$ tensor in orbital space. The generalized response is defined by the orbital-resolved correlation functions 
\begin{align}
\Pi_{ij}^{\mu\nu\mu'\nu'}({\bf q}, i\omega_n)
  &= \frac{1}{N} \int_0^{\beta} d\tau\, e^{i \omega_n \tau} \nonumber \\
&{}\times
  \left\langle T_\tau\, \sigma_i^{\mu\nu} ({\bf q},\tau)\, \sigma_j^{\mu'\nu'}(-{\bf q},0) \right\rangle ,
\label{response_general}
\end{align}
where
\begin{equation}
\sigma_l^{\mu\nu} = \sum_{{\bf k},\zeta,\zeta'} c_{{\bf k} + {\bf q},\mu,\zeta}^\dagger\,
  \sigma^l_{\zeta,\zeta'}\, c_{{\bf k},\nu,\zeta'}
\end{equation}
is the Fourier-transformed spin operator, given in terms of the fermionic annihilation (creation) operators $c_{{\bf k},\mu,\sigma}$ ($c_{{\bf k},\mu,\sigma}^\dagger$) for orbital $\mu$. $\beta=1/k_BT$ is the inverse temperature, $i\omega_n$ are Matsubara frequencies, and $T_\tau$ is the time-ordering directive in imaginary time. Here, the index $l = 0$ signifies the density and $l = x,y,z$ refers to the spin components. The retarded response functions are obtained by the analytic continuation $i\omega_n \to \omega + i\delta$.

\subsection{Bulk response}

The zero-temperature dynamical density and spin response of the bulk states has been calculated in Ref.\ \cite{Ghosh2019,Zhou2018} using a ${\bf k}\cdot{\bf p}$ Hamiltonian. For the sake of completeness and to explore its orbital structure, we calculate the bulk response tensor for our model. The details of the evaluation are relegated to Appendix~\ref{app_bulk_response}. As noted, each component of the density-spin response tensor is a $2 \times 2 \times 2 \times 2$ tensor in the orbital basis. The terms can be divided into four intra-orbital terms $\Pi_{ij}^{\mu\mu\nu\nu}$, four inter-orbital terms $\Pi_{ij}^{\mu\bar{\mu}\mu\bar{\mu}}$ and $\Pi_{ij}^{\mu\bar{\mu}\bar{\mu}\mu}$, where $\bar{\mu}$ is the orbital other than $\mu$, and eight terms for which not all superscripts appear in pairs. We find that for the pure density-density and spin-spin response, the orbital structure is trivial, i.e., all intra-orbital and inter-orbital terms are equal, while the other terms vanish. On the other hand, the intra-orbital and inter-orbital contributions to the coupled density-spin response have opposite signs for states near the two Weyl nodes of opposite chirality and therefore vanish in equilibrium. However, terms such as $\Pi_{0l}^{\mu\mu\mu\bar{\mu}}$ survive for the coupled density-spin response. Although such terms are not physically relevant at the non-interacting level because they do not enter the coupled density-spin response, they can influence the current response for interacting electrons, as discussed below. The rather lengthy expressions for bulk response functions for our model are given in Appendix\ \ref{app_bulk_response}. They are similar to the ones presented in Ref.~\cite{Ghosh2019}.

\subsection{Fermi-arc response}
\label{sub.1FA}

Next, we investigate the zero-temperature dynamical response of the FA surface states \footnote{The cross-correlations between FA and bulk states vanish since the FA states are eigenspinors of $\sigma_x \otimes \tau_x$, whereas the bulk states are eigenspinors of $\sigma_0 \otimes \tau_x$.}. The orbital components of the response tensor can be written as
\begin{align}
&\Pi_{ij}^{\mu\nu\mu'\nu'}({\bf q}, i\omega_n) = - \frac{1}{N}
  \sum_{{\bf k}_\parallel} \int dz \int dz' \nonumber \\
&\quad{}\times \langle \phi^{\mu}({\bf k}_\parallel + {\bf q},z)|
    \sigma_i |\phi^{\nu}({\bf k}_\parallel,z)\rangle \nonumber \\
&\quad{}\times \langle \phi^{\mu'}({\bf k}_\parallel,z')|
    \sigma_j |\phi^{\nu'}({\bf k}_\parallel + {\bf q},z')\rangle \nonumber \\
&\quad{}\times \frac{n_F({\bf k}_\parallel)
    - n_F({\bf k}_\parallel + {\bf q})}
  {i\omega_n + \epsilon({\bf k}_\parallel)
    - \epsilon({\bf k}_\parallel + {\bf q})} ,
\label{response_FA}
\end{align}
where $n_F({\bf k}_\parallel)$ is the Fermi function for the FA states with two-component wave vector ${\bf k}_\parallel$, ${\bf q} = (q_x,q_y)$ is the wave vector of the response, and $\phi^{\mu}({\bf k}_\parallel,z)$ is the component of the FA wave function for orbital $\mu$, see Eqs.\ (\ref{wave_function_FA.1}) and (\ref{wave_function_FA.2}). The integrals over $z$ and $z'$ take into account the extension of FA states into the bulk and therefore run from $-\infty$ to $0$ for the top surface and $0$ to $\infty$ for the bottom surface, in the limit of infinite thickness. This is important because it allows for the long tail of the FA states near the projection of the Weyl nodes. A purely two-dimensional response function calculated by projecting the FA states into the $k_xk_y$-plane would yield qualitatively incorrect result, even though the FA states are surface states energetically separated from the bulk. 

Redefining the orbital basis as $\{|{\oplus}{\oplus}\rangle$, $|{\oplus}{\ominus}\rangle$, $|{\ominus}{\oplus}\rangle$, ${|\ominus}{\ominus}\rangle\}$, the $2 \times 2 \times 2 \times 2$ blocks in the response tensor can be written as $4 \times 4$ matrices. We find that all components of the composite density-spin response tensor in the orbital basis can be expressed by four distinct terms $\Pi_1$, $\Pi_2$, $\Pi_3$, and $\Pi_4$, which are given by
\begin{widetext}
\begin{align}
\Pi_1^{T,B} &= \frac{1}{(2\pi)^2}\, \frac{q_x}{2\lambda q_x \pm (\omega + i \delta)} \, \frac{1}{8} \left[\frac{2b_y}{\lambda} -4 |q_y| + \frac{\lambda q_y^2}{b_y} + \frac{\lambda^2 |q_y|^3}{b_y^2} + \frac{\lambda^3 q_y^4}{b_y^3} \arctanh \left(1-\frac{\lambda |q_y|}{b_y}\right) \right],
\label{Pi_TB.1} \\
\Pi_2^{T,B} &= \frac{1}{(2\pi)^2}\, \frac{q_x}{2\lambda q_x \pm (\omega + i \delta)} \, \frac{1}{24b_y^3 \lambda}
\nonumber \\
&{}\times \bigg[ b_y \left(b_y - \lambda |q_y|\right) \left( 2b_y^2 - 10 \lambda |q_y| b_y - 13 \lambda^2 q_y^2 \right) + 3\lambda^2 q_y^2 \left(8b_y^2 - \lambda^2 q_y^2\right) \arctanh \left(1-\frac{\lambda |q_y|}{b_y}\right) \bigg],
\label{Pi_TB.2} \\
\Pi_3^{T,B} &= \frac{1}{(2\pi)^2}\, \frac{q_x}{2\lambda q_x \pm (\omega + i \delta)} \, \frac{1}{4(b_y^2 - \lambda^2 q_y^2 )^2}\, \frac{8}{15\lambda} \left(b_y - \lambda |q_y|\right)^3 \left(b_y^2 + 3b_y \lambda |q_y| + \lambda^2 q_y^2\right), \\
\Pi_4^{T,B} &= - \frac{1}{(2\pi)^2}\, \frac{q_x}{2\lambda q_x \pm (\omega + i \delta)} \, \frac{1}{4(b_y^2 - \lambda^2 q_y^2 )} \left[ \frac{2b_y^3 }{3\lambda} + b_y \lambda q_y^2 - \frac{5}{3} \lambda^2 |q_y|^3 + \left(\frac{\lambda^3 q_y^4}{b_y} - 4b_y \lambda q_y^2\right) \arctanh \left(1-\frac{\lambda |q_y|}{b_y}\right)\right].
\end{align}
\end{widetext}
Here, the superscripts $T$, $B$ as well as the signs $+$, $-$ refer to the top and bottom surface, respectively. It should be noted that the $\Pi_\nu^{T,B}$ are functions of the wave vector ${\bf q}$ and the frequency $\omega$ but we will suppress these arguments from now on for brevity. The density response is now given by
\begin{equation}
\Pi^{T,B}_{00}
= \left( \arraycolsep=1.6pt\def\arraystretch{1.5}
\begin{array}{cccc}
\Pi^{T,B}_1 & 0 & 0 & \Pi^{T,B}_1 \\
0 & \Pi^{T,B}_2 & \Pi^{T,B}_2 & 0 \\
0 & \Pi^{T,B}_2 & \Pi^{T,B}_2 & 0 \\
\Pi^{T,B}_1 & 0 & 0 & \Pi^{T,B}_1  
\end{array}\right) .
\label{Pi00}
\end{equation} 
For illustration, we describe the evaluation of this matrix in Appendix\ \ref{app_Pi00}. Evidently, the density response only contains intra-orbital (${\oplus}{\oplus}{\oplus}{\oplus}$, ${\oplus}{\oplus}{\ominus}{\ominus}$, ${\ominus}{\ominus}{\oplus}{\oplus}$, ${\ominus}{\ominus}{\ominus}{\ominus}$) and inter-orbital (${\oplus}{\ominus}{\oplus}{\ominus}$, ${\ominus}{\oplus}{\ominus}{\oplus}$, ${\oplus}{\ominus}{\ominus}{\oplus}$, ${\ominus}{\oplus}{\oplus}{\ominus}$) contributions. A few additional remarks are in order. Equations (\ref{Pi_TB.1}) and (\ref{Pi_TB.2}) show that the density response is even in $q_y$, which results from the FA dispersion being symmetric under $q_y\to -q_y$. However, the response tensor is nonanalytic at $q_y=0$. The origin is that the surface bands are restricted to $-b_y/2\lambda \le k_y \le b_y/2\lambda$, which means that the $k_y$ integral contained in Eq.\ (\ref{response_FA}) is limited to an interval of length $b_y/\lambda - |q_y|$. One might suspect this nonanalyticity to be an artifact of truncating the surface band where it really merges into the bulk states. We suggest that this is not the case, based on the following reasoning: for all $q_y\neq 0$, the contribution to the response of FA states close to the ends of the arcs, i.e., for $k_y\to \pm b_y/2\lambda$, approaches zero. This is due to the destructive interference between these very extended states for different $k_y$. However, the point $q_y=0$ is special: the matrix elements in Eq.\ (\ref{response_FA}) are then essentially normalization integrals so that all $k_y \in (-b_y/2\lambda, b_y/2\lambda)$ contribute equally, regardless of the decay length. Although the corresponding discontinuities at $k_y=\pm b_y/2\lambda$ only exist for $q_y=0$, they are sufficient to generate odd powers of $|q_y|$ in the response functions. Inclusion of the bulk response does not cure this nonanalyticity since, in the thermodynamic limit, the missing spectral weight in the bulk resulting from the formation of surface states has negligible effect on the response. Moreover, the response satisfies $\Pi^{T,B}_{00}(-{\bf q},-\omega) = \Pi^{T,B}_{00}({\bf q},\omega)^*$, which follows from the definition (\ref{response_FA}). Unlike the density response of a two-dimensional electron gas, there is no symmetry under inversion of ${\bf q}$ alone. This reflects the chiral nature of the FAs.

Similarly, the spin response is found to be
\begin{align}
\Pi^{T,B}_{xx} 
&= \left( \arraycolsep=1.6pt\def\arraystretch{1.5} \begin{array}{cccc}
\Pi^{T,B}_2 & 0 & 0 & \Pi^{T,B}_2 \\
0 & \Pi^{T,B}_1 & \Pi^{T,B}_1 & 0 \\
0 & \Pi^{T,B}_1 & \Pi^{T,B}_1 & 0 \\
\Pi^{T,B}_2 & 0 & 0 & \Pi^{T,B}_2  
\end{array}\right) ,
\label{Pixx} \\
\Pi^{T,B}_{yy} 
&= \Pi^{T,B}_3 \left( \arraycolsep=1.6pt\def\arraystretch{1.5} \begin{array}{cccc}
1 & -p & p & -1 \\
p & -p^2 & p^2 & -p \\
-p & p^2 & -p^2 & p \\
-1 & p & -p & 1
\end{array}\right) ,
\label{Piyy} \\
\Pi^{T,B}_{zz} 
&= \Pi^{T,B}_3 \left( \arraycolsep=1.6pt\def\arraystretch{1.5} \begin{array}{cccc}
p^2 & -p & p & -p^2 \\
p & -1 & 1 & -p \\
-p & 1 & -1 & p \\
-p^2 & p & -p & p^2
\end{array}\right) ,
\label{Pizz} \\
\Pi^{T,B}_{xy} 
&= \pm \Pi^{T,B}_4 \left( \arraycolsep=3.2pt\def\arraystretch{1.5} \begin{array}{cccc}
0 & 0 & 0 & 0 \\
-1 & p & -p & 1 \\
-1 & p & -p & 1 \\
0 & 0 & 0 & 0
\end{array}\right)  ,
\label{Pixy} \\
\Pi^{T,B}_{yx} 
&= \pm \Pi^{T,B}_4 \left( \arraycolsep=3.2pt\def\arraystretch{1.5} \begin{array}{cccc}
0 & -1 & -1 & 0 \\
0 & -p & -p & 0 \\
0 & p & p & 0 \\
0 & 1 & 1 & 0
\end{array}\right) ,
\label{Piyx} \\
\Pi^{T,B}_{yz} 
&= \pm i \Pi^{T,B}_3 \left( \arraycolsep=1.6pt\def\arraystretch{1.5} \begin{array}{cccc}
p & -1 & 1 & -p \\
p^2 & -p & p & -p^2 \\
-p^2 & p & -p & p^2 \\
-p & 1 & -1 & p
\end{array}\right) ,
\label{Piyz} \\
\Pi^{T,B}_{zy} 
&= \pm i \Pi^{T,B}_3 \left( \arraycolsep=1.6pt\def\arraystretch{1.5} \begin{array}{cccc}
-p & p^2 & -p^2 & p \\
-1 & p & -p & 1 \\
1 & -p & p & -1 \\
p & -p^2 & p^2 & -p
\end{array}\right) ,
\label{Pizy} \\
\Pi^{T,B}_{zx} 
&= i \Pi^{T,B}_4 \left( \arraycolsep=3.2pt\def\arraystretch{1.5} \begin{array}{cccc}
0 & p & p & 0 \\
0 & 1 & 1 & 0 \\
0 & -1 & -1 & 0 \\
0 & -p & -p & 0
\end{array}\right) ,
\label{Pizx} \\
\Pi^{T,B}_{xz} 
&= i \Pi^{T,B}_4 \left( \arraycolsep=3.2pt\def\arraystretch{1.5} \begin{array}{cccc}
0 & 0 & 0 & 0 \\
-p & 1 & -1 & p \\
-p & 1 & -1 & p \\
0 & 0 & 0 & 0
\end{array}\right) ,
\label{Pixz}
\end{align}
where $p = \lambda q_y /b_y$.

The coupled density-spin response is described by
\begin{align}
\Pi^{T,B}_{0x} &= \left( \arraycolsep=1.6pt\def\arraystretch{1.5} \begin{array}{cccc}
0 & \Pi^{T,B}_1 & \Pi^{T,B}_1 & 0 \\
\Pi^{T,B}_2 & 0 & 0 & \Pi^{T,B}_2 \\
\Pi^{T,B}_2 & 0 & 0 & \Pi^{T,B}_2 \\
0 & \Pi^{T,B}_1 & \Pi^{T,B}_1 & 0  
\end{array}\right) ,
\label{Pi0x} \\
\Pi^{T,B}_{x0} &= \left( \arraycolsep=1.6pt\def\arraystretch{1.5} \begin{array}{cccc}
0 & \Pi^{T,B}_2 & \Pi^{T,B}_2 & 0 \\
\Pi^{T,B}_1 & 0 & 0 & \Pi^{T,B}_1 \\
\Pi^{T,B}_1 & 0 & 0 & \Pi^{T,B}_1 \\
0 & \Pi^{T,B}_2 & \Pi^{T,B}_2 & 0  
\end{array}\right) ,
\label{Pixo} \\
\Pi^{T,B}_{0y} 
&= \Pi^{T,B}_4 \left( \arraycolsep=3.2pt\def\arraystretch{1.5} \begin{array}{cccc}
1 & -p & p & -1 \\
0 & 0 & 0 & 0 \\
0 & 0 & 0 & 0 \\
1 & -p & p & -1
\end{array}\right) ,
\label{Pi0y} \\
\Pi^{T,B}_{y0} 
&= \Pi^{T,B}_4 \left( \arraycolsep=3.2pt\def\arraystretch{1.5} \begin{array}{cccc}
1 & 0 & 0 & 1 \\
p & 0 & 0 & p \\
-p & 0 & 0 & -p \\
-1 & 0 & 0 & -1
\end{array}\right) ,
\label{Piy0} \\
\Pi^{T,B}_{0z}
&= \pm i \Pi_4^{T,B} \left( \arraycolsep=3.2pt\def\arraystretch{1.5} \begin{array}{cccc}
 p & -1 & 1 & -p \\
0 & 0 & 0 & 0 \\
0 & 0 & 0 & 0 \\
p & -1 & 1 & -p  
\end{array}\right) ,
\label{Pi0z} \\
\Pi^{T,B}_{z0}
&= \pm i \Pi_4^{T,B} \left( \arraycolsep=3.2pt\def\arraystretch{1.5} \begin{array}{cccc}
 -p & 0 & 0 & -p \\
-1 & 0 & 0 & -1 \\
1 & 0 & 0 & 1 \\
p & 0 & 0 & p  
\end{array}\right) .
\label{Piz0}
\end{align}
The response functions do not depend on the electron filling or chemical potential since the FA density of states is independent of energy. To leading order for small wave vectors, $\lambda q \ll \omega, b_y$, the real part of the contributions reads
\begin{align}
{\rm Re}\: \Pi_1^{T,B} &= \pm \frac{1}{(2\pi)^2}\, \frac{q_x}{\omega}
  \left(1 \mp \frac{2\lambda q_x}{\omega}\right) \frac{1}{4\lambda}\, (b_y - 2\lambda q_y), \\
{\rm Re}\: \Pi_2^{T,B} &= \pm \frac{1}{(2\pi)^2}\, \frac{q_x}{\omega}
  \left(1 \mp \frac{2\lambda q_x}{\omega}\right) \frac{1}{12\lambda}\, (b_y - 6\lambda q_y), \\
{\rm Re}\: \Pi_3^{T,B} &= \pm \frac{1}{(2\pi)^2}\, \frac{q_x}{\omega}
  \left(1 \mp \frac{2\lambda q_x}{\omega}\right) \frac{2}{15\lambda}\, b_y, \\
{\rm Re}\: \Pi_4^{T,B} &= \pm \frac{1}{(2\pi)^2}\, \frac{q_x}{\omega}
  \left(1 \mp \frac{2\lambda q_x}{\omega}\right) \frac{1}{6\lambda}\, b_y.
\end{align}
Therefore, similarly to the results of Ref.~\cite{Andolina2018} for a WSM described by a ${\bf k}\cdot{\bf p}$ Hamiltonian, the density response of FA states in our model varies as $\sim q_x/\omega$ to lowest order. Neglecting the tail of the FA states by using a purely two-dimensional model results in an erroneous $1/\omega^2$ dependence~\cite{Losic2018}.

\section{Related physical quantities and observable consequences}
\label{sec4}

In this section, we will analyze physical effects that are determined by the response functions calculated in Sec.\ \ref{sec3}. Specifically, we will address the bulk and surface contributions to the electromagnetic response and the optical conductivity.

\subsection{Bulk response}

Here, we discuss effects of the bulk of WSMs. Some of these have been considered in Ref.~\cite{Zhou2018} but are included here for completeness as well as to address the orbital structure of the response.

\subsubsection{Electromagnetic susceptibilities}

The density-spin response tensor physically manifests itself by the electromagnetic susceptibilities. The electric susceptibility can be expressed in terms of only the intra-orbital density-density correlation function as
\begin{equation}
\chi^{({e})\mu \nu}_{ij} = \frac{\partial P^{\mu}_i}{\partial E^{\nu}_j}
  = - \frac{e^2}{q_i q_j}\, \Pi^{\mu \mu \nu \nu}_{00},
\end{equation}
while the magnetic (spin) susceptibility is given by the spin-spin correlations as 
\begin{equation}
\chi^{({m})\mu \nu}_{ij} = \frac{\partial M^{\mu}_i}{\partial B^{\nu}_j}
  = \left( \frac{\mu_B}{2} \right)^{\!2} g^{\mu} g^{\nu}\, \Pi^{\mu \mu \nu \nu}_{ij},
\end{equation}
where $\mu_{B}$ is the Bohr magneton and $g^{\mu}$ the $g$-factor for orbital $\mu$. As the two orbitals couple to the electric field with the same electronic charge $-e$, the net polarization response to a electric field for bulk states can be calculated by tracing over orbital indices of the density response, $\sum_{\mu,\nu} \chi^{({e})\mu \nu}_{00}$. On the other hand, since the two orbitals can have different $g$-factors \footnote{The two orbitals have opposite parity and are thus expected to have different $g$-factors since the $g$-factors originate from the complex interplay of chemical bonding, crystal-field splitting, and spin-orbit coupling.}, application of an external magnetic field generally results in different spin polarizations for the two orbitals. 

The spin susceptibility can be expressed in terms of longitudinal and transverse components,
\begin{equation}
\chi^{({m})\mu \nu}_{ij}({\bf q})
  = \chi^{({m})\mu \nu}_{L}({\bf q})\, \frac{q_i q_j}{q^2}
  + \chi^{({m})\mu \nu}_{T}({\bf q}) \left( \delta_{ij} - \frac{q_i q_j}{q^2} \right) .
\end{equation} 
The Pauli spin susceptibility, defined as the static longitudinal spin susceptibility for ${\bf q}\rightarrow 0$, is found to vanish identically \cite{Koshino2016}. This can be attributed to the spin-momentum locking in the Weyl Hamiltonian near the nodes and has recently been observed in the candidate materials NbP and TaP \cite{Leahy2018}. The vanishing Pauli susceptibility holds true even beyond the linear-response regime~\cite{Koshino2016}. 

The crossed magnetoelectric susceptibility given by the coupled density-spin response vanishes in equilibrium in our model. However, in the presence of non-orthogonal static ${\bf E}$ and ${\bf B}$ fields, the two nodes develop different effective chemical potentials due to the chiral anomaly, which leads to a nonzero magnetoelectric response~\cite{Ghosh2019}.

\subsubsection{Current response}

Now we look at the current-current correlations, which govern the response of a system to the vector potential. The current operator corresponding to the Weyl Hamiltonian in Eq.\ (\ref{hamil2}) reads $\hat{\jmath}_l = 2\lambda_l e\, \sigma_l \otimes \tau_x$. Consequently, the current-current correlations are related to the spin-spin correlations as
\begin{equation}
\Pi_{j_l j_m} = 4 \lambda_l \lambda_m e^2 \sum_{\mu} \big(\Pi^{\mu \bar{\mu} \mu \bar{\mu}}_{lm}
  + \Pi^{\mu \bar{\mu} \bar{\mu} \mu}_{lm}\big)
\end{equation}
and thus depend only on the inter-orbital response. The longitudinal and transverse current response is proportional to the longitudinal and transverse spin response, respectively. 

The optical conductivity of WSM can be obtained from the current-current correlation function using
\begin{equation}
\sigma_{lm}(\omega) = -\frac{i}{\omega + i \delta}\, \lim_{{\bf q}\rightarrow 0} \Pi_{j_l j_m}(\omega),
\end{equation}
which in our case yields
\begin{equation}
{\rm Re}\: \sigma_{lm}(\omega) =  \left( \frac{e^2 \mu^2}{3\pi v_F}\, \delta(\omega)
  + \frac{e^2 \omega}{3\pi v_F}\, \Theta(\omega - 2 \mu)\right) \delta_{lm},
\end{equation}
where $\mu$ is the chemical potential, $v_F$ is the Fermi velocity, and $\Theta(x)$ is the Heaviside step function. Here, we have taken $\omega,\mu \geq 0$. Thus, outside the Pauli-blockade regime, i.e., for $\omega > 2 \mu$, the interband optical conductivity for free Weyl fermions is linear in $\omega$ which is consistent with previous results \cite{Hosur2012,Ashby2014}.

A related observable is the orbital magnetic susceptibility which is proportional to the transverse current-current correlation function \cite{Giuliani2005}. Its evaluation reveals that the orbital susceptibility is diamagnetic and varies as $\sim \ln \mu$. This logarithmic dependence on the chemical potential is consistent with more rigorous calculations invoking Landau-level quantization \cite{Koshino2010,Koshino2016,Mikitik2019}.

\subsection{Fermi-arc response}

In this paper, we are mainly interested in the effects of the FA surface states. In the following, observables related to the FA response are analyzed.

\subsubsection{Electromagnetic susceptibilities}
\label{subsubsec_sus}

Similarly to the bulk states, the electric susceptibility of the FA states is related to the intra-orbital density response, $\chi^{(e)}_{ij}= -({e^2}/{q_i q_j})\, \Pi^{T,B}_1$. Also, the spin susceptibilities are given by the spin-spin response as $\chi^{({m})\mu \nu}_{ij} = {\partial M^{\mu}_i}/{\partial B^{\nu}_j} = (\mu_B/2)^2\, g^{\mu} g^{\nu}\, \Pi^{\mu\mu\nu\nu}_{ij}$. Since the FA states are eigenstates of $\sigma_x$, one may naively assume that the spin susceptibilities for $y$ and $z$ spin components vanish. However, this is not generally true. The spin susceptibilities
\begin{align}
\chi^{({m}){T,B}}_{yy} &= \left( \frac{\mu_B}{2} \right)^{2} (g^{\oplus} - g^{\ominus})^2 \ \Pi_{3}^{T,B}, \\\chi^{({m}){T,B}}_{zz} &= \left( \frac{\mu_B}{2} \right)^{2} (g^{\oplus} - g^{\ominus})^2 p^2 \ \Pi_{3}^{T,B},\\
\chi^{({m}){T,B}}_{yz} &= \left( \frac{\mu_B}{2} \right)^{2} (g^{\oplus} - g^{\ominus})^2 p \ \Pi_{3}^{T,B}
\end{align}
will have non-zero values owing to different $g$ factors for the two orbitals.
 
The more important manifestation of the FA response appears in the magnetoelectric effect. The mixed magnetoelectric susceptibilities are given by the crossed density-spin response functions as
\begin{align}
\chi^{({em})\mu\nu}_{ij} &= \frac{\partial P^{\mu}_i}{\partial B^{\nu}_j}
  = i\, \frac{e \mu_B}{2}\, \frac{1}{q_i}\, g^{\nu}\, \Pi^{\mu \mu \nu \nu}_{0j} ,
\label{eq_sus_electromagnetic}  
  \\
\chi^{({me})\mu\nu}_{ij} &= \frac{\partial M^{\mu}_i}{\partial E^{\nu}_j}
  = i\, \frac{e \mu_B}{2}\, \frac{1}{q_j}\, g^{\mu}\, \Pi^{\mu \mu \nu \nu}_{i0}.
\label{eq_sus_magnetoelectric} 
\end{align}
These relations show that the FA states exhibit the magnetoelectric effect. A closer look reveals that a static, homogeneous magnetic field along the growth direction ($z$) yields a charge polarization along the node-separation axis ($y$),
\begin{equation}
P_y = \pm \frac{1}{(2\pi)^2}\, \frac{e \mu_B}{12 \lambda}\, (g^\oplus - g^\ominus) B_z,
\end{equation}    
while an electric field along the node-separation direction results in a magnetization along the growth axis,
\begin{equation}
M_z = \mp \frac{1}{(2\pi)^2}\, \frac{e \mu_B}{12 \lambda}\, (g^\oplus - g^\ominus) E_y.
\end{equation}
The upper (lower) sign refers to the top (bottom) surface, indicating the chiral origin of the response. This is one of the key findings of this work. Using typical experimental values for Fermi velocity and the separation of the Weyl nodes, see Sec.\ \ref{secVB1}, and assuming $g^\oplus - g^\ominus \approx 1$, the magnetoelectric coefficient is estimated to be 10$^{-31}\, \mathrm{s/m}$ per fermion.

The bulk WSM states also exhibit a magnetoelectric effect \cite{Ghosh2019}, albeit not in the static limit. Importantly, the bulk magnetoelectric effect only occurs in nonequilibrium situations in the presence of non-orthogonal static ${\bf E}$ and ${\bf B}$ fields. In contrast, the surface magnetoelectric effect survives in equilibrium and can thus serve as a hallmark of the FA states.

Finally, the Pauli spin susceptibility of the FA states reads
\begin{align}
\chi_{\rm Pauli} &= \sum_{\mu,\nu} \lim_{{\bf q} \rightarrow 0} \chi^{({m}),\mu \nu}_{xx}({\bf q},0) \nonumber \\
&= \left(\frac{\mu_B}{2}\right)^2\, (g^\oplus + g^\ominus)^2\, \frac{b_y}{24 \lambda^2},
\end{align}
which is proportional to the FA density of states $\mathcal{N}^{T,B} = b_y/8\pi^2 \lambda^2$, similarly to the conventional electron gas.

\subsubsection{Current response and anomalous Hall effect}

Next, we study the current response of the FA states. Since these states only disperse along $k_x$, only the $x$-component of the current operator, $\hat{\jmath}_x = 2\lambda e\, \sigma_x \otimes \tau_x$, is nonzero. We see from Eqs.\ (\ref{wave_function_FA.1}) and (\ref{wave_function_FA.2}) that the FA states are eigenstates of $\hat{\jmath}_x$ and thus carry a finite surface current, analogous to the persistent currents for quantum Hall edge states. This analogy stems from the fact that for $- b_y/2\lambda < k_y < b_y/2\lambda$, each two-dimensional Hamiltonian $\mathcal{H}_{k_y}(k_x,k_z)$ in the $k_xk_z$-plane represents a two-dimensional Chern insulator. The FAs are the chiral edge states of the Chern insulators and therefore exhibit the quantum anomalous Hall effect {\cite{Burkov2011,Armitage2018}.

We now calculate the surface current carried by the FAs. The FA states are eigenstates of $\hat{\jmath}_x$ with opposite eigenvalues for the top and bottom surfaces. In equilibrium, the currents contributed by the top and bottom surfaces are equal and opposite, leading to a vanishing total current. The application of a voltage $V_H$ along $z$ leads to a difference between the chemical potentials for the two surfaces $\Delta\mu = e V_H$. In this case, we obtain a total current from the population imbalance of the chiral edge states,
\begin{align}
&I_x = e V_H \left|\langle \hat{\jmath}_x \rangle^{T}
  - \langle \hat{\jmath}_x \rangle^{B}\right| \nonumber \\
&= e V_H\, \frac{2\lambda e}{(2\pi)^2}\,
  \Bigg( \int dk_x \int_{-b_y/2\lambda}^{b_y/2\lambda} dk_y \int_0^\infty dz\,
  |\Phi^{B}({\bf k}_\parallel,z)|^2 \nonumber \\
&\quad{}- \int dk_x \int_{-b_y/2\lambda}^{b_y/2\lambda} dk_y \int_{-\infty}^0 dz\,
  |\Phi^{T}({\bf k}_\parallel,z)|^2 \Bigg) \nonumber \\
&= \frac{e^2}{2\pi}\, \frac{b_y}{\lambda}\, V_H .
\end{align}
Therefore, the anomalous Hall conductivity is
\begin{equation}
\sigma_{xz} = \frac{e^2}{2\pi}\, \frac{b_y}{\lambda} ,
\end{equation}
which is equal to the distance between the Weyl nodes in units of $e^2/2\pi$~\cite{Burkov2011,Burkov2014}.

The current-current correlation function reads
\begin{equation}
\Pi^{T,B}_{j_x j_x} = 4\lambda^2 e^2 \sum_{\mu = \oplus,\ominus} \big(\Pi^{\mu\bar{\mu}\mu\bar{\mu}}_{xx}
  + \Pi^{\mu\bar{\mu}\bar{\mu}\mu}_{xx}\big) = 16 \lambda^2 e^2\, \Pi^{T,B}_1
\end{equation}
and is thus proportional to the density susceptibility. It is important to note that the density and current response of the bulk states are related by a Ward identity \cite{Ward1950} since the bulk carrier density and current obey a continuity equation. There is no corresponding identity for the FA states since the total carrier density at the surfaces consists of contributions from FA and bulk states, and it is the total carrier density that is conserved. There is no continuity equation satisfied by FA carriers separately.

\section{Response of the interacting system}
\label{sec5}

\begin{figure}[tb]
\includegraphics[width=1.0\columnwidth]{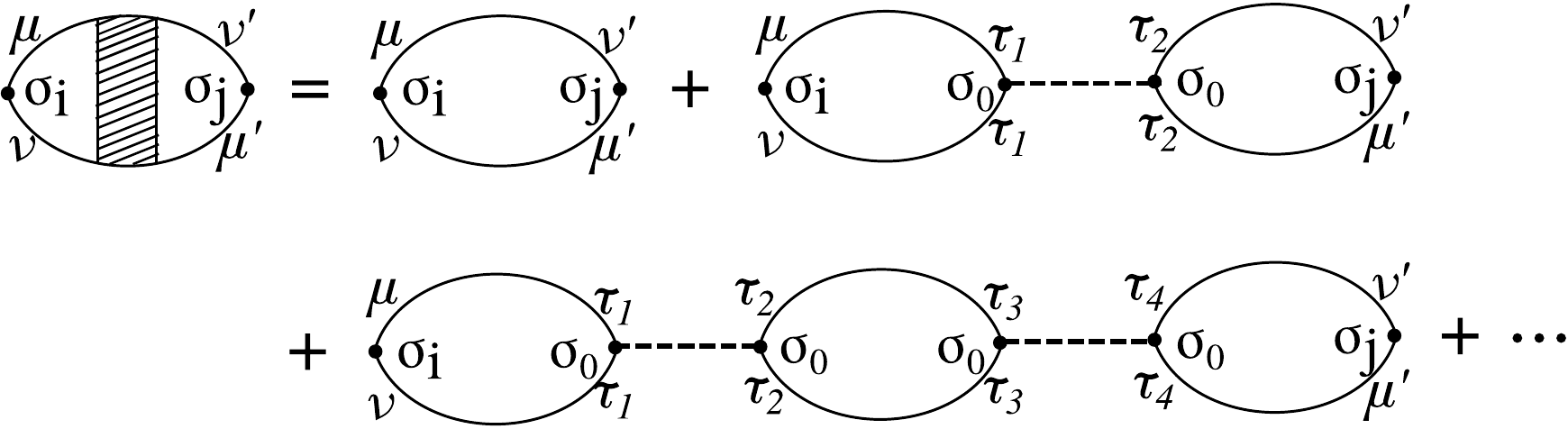}
\caption{Diagrammatic representation of the density-spin response function in the RPA. The dashed line corresponds to the bare Coulomb interaction $V_q$.}
\label{fig_RPA} 
\end{figure} 

Let us now consider the response functions for interacting electrons. The response functions are calculated within the RPA, depicted diagrammatically in Fig.\ \ref{fig_RPA}. Although the RPA is a weak-coupling approximation, it provides a quantitatively accurate description of WSMs even in the strong-coupling regime \cite{Hofmann2014,Hofmann2015}. The RPA is thus justified for our calculations of the interacting response functions. Evaluating the perturbative expansion in orders of the Coulomb interaction $V_q$, the interacting response function can be expressed as
\begin{align}
\widetilde{\Pi}_{ij}^{\mu\nu\mu'\nu'}({\bf q}, \omega)
  &= \Pi_{ij}^{\mu\nu\mu'\nu'}({\bf q}, \omega) \nonumber\\
&{}- \frac{\Pi_{i0}^{\mu\nu\gamma\gamma}({\bf q}, \omega)\, V_q\,
  \Pi_{0j}^{\delta\delta\mu'\nu'}({\bf q}, \omega) }
  {1 + V_q\, \Pi_{00}^{\alpha\alpha\beta\beta}({\bf q}, \omega)},
\label{RPA}
\end{align}
where summation over repeated indices is implied.

\subsection{Interacting bulk response}

First, we study the response of interacting electrons in bulk WSM states. Here, $V_q = 4 \pi e^2 / \kappa q^2$ is the Fourier-transformed Coulomb interaction, with $\kappa$ being the dielectric constant. After algebraic manipulations of Eq.\ (\ref{RPA}), the physical density response reads
\begin{equation}
 \widetilde{\Pi}_{00}({\bf q}, \omega)
  =  \sum_{\mu,\nu} \widetilde{\Pi}^{\mu \mu \nu \nu}_{00}({\bf q}, \omega)
  =  \frac{\Pi_{00} ({\bf q}, \omega)}{1 + V_q\, \Pi_{00} ({\bf q}, \omega)}.
\end{equation}
Clearly, the density response is enhanced by a factor $1/(1 + V_q\, \Pi_{00})$, which is the inverse RPA dielectric function. The zeros of the RPA dielectric function correspond to collective density excitations, i.e., plasmons. The plasmon dispersion can be obtained in the long-wavelength limit, keeping only the leading order in ${\bf q}$, as carried out in Refs.\ \cite{Lv2013,Ghosh2019}, with the result
\begin{equation}
\omega_{\rm pl}({\bf q}) = \omega_0 \left( 1 - \frac{({\bf v}_F \cdot {\bf q})^2}{8 \mu^2}
  \left[ 1 +  \frac{\nu_0^2 - 3/5}{\nu_0^2\, (1-\nu_0^2)^2} \right] \right) ,
\end{equation}
where
\begin{equation}
\omega_0 = \mu\, \sqrt{\frac{16 \alpha_{\kappa}}{3 \pi \kappa^*(\omega_0)}}
\end{equation}
is the plasma frequency at $\mathbf{q}\rightarrow 0$, $\nu_0 = \omega_0/2\mu$, and ${\bf v}_F$ is the vectorial Fermi velocity. The ${\bf q}\to 0$ plasma frequency is determined by the effective fine structure constant $\alpha_{\kappa} = e^2/\kappa v_F$ and the frequency-dependent effective background dielectric function
\begin{equation}
\kappa^*(\omega) = 1 + \frac{4\alpha_{\kappa}}{3\pi}\,
  \ln \left| \frac{4 \varepsilon_c^2}{4 \mu^2 - \omega^2} \right| .
\end{equation}
Thus, the plasmon dispersion is gapped and the leading momentum dependence is quadratic. The dispersion is manifested as sharp peaks in the electron energy-loss function, which is experimentally accessible. In the presence of non-orthogonal ${\bf E}$ and ${\bf B}$ fields, the plasmons carry spin, which can be probed by optical pump-probe spectroscopy as discussed in Ref.~\cite{Ghosh2019}.

On the other hand, Eq.\ (\ref{RPA}) shows that in equilibrium the physical spin response and consequently the spin susceptibilities are not affected by the electron interaction at the RPA level. This is because the crossed density-spin functions vanish in equilibrium, as discussed above. The current-current response, however, is influenced by the Coulomb interaction and reads 
\begin{equation}
\widetilde{\Pi}_{j_l j_m}({\bf q}, \omega) = \frac{\Pi_{j_l j_m}({\bf q}, \omega)}
  {1 + V_q\, \Pi_{00} ({\bf q}, \omega)}.
\end{equation}
Thus, the current-current correlation function is renormalized by the RPA dielectric functions. As a consequence, the interband optical conductivity is no longer linear in $\omega$ but rather shows a more complicated $\omega$ dependence given by
\begin{equation}
{\rm Re}\, \sigma_{lm}(\omega)
  = \frac{e^2 }{3\pi v_F}\,
  \frac{\omega}{\kappa^*(\omega) - \frac{16 \mu^2 \alpha_{\kappa}}{3 \pi \omega^2}}\,
  \Theta(\omega -2\mu)\, \delta_{lm}.
\end{equation}

Similarly, the orbital magnetic susceptibility is also strongly influenced by the Coulomb interaction. Not only its logarithmic dependence on the chemical potential is affected, a transition to orbital paramagnetism, signalled by a sign change of the orbital magnetic susceptibility, could occur, as discussed in Ref.~\cite{Zhou2018}. 

\subsection{Interacting Fermi-arc response}

Second, we study the effect of electron interactions on the dynamical response of the FA states. We first address the density response and the FA plasmons and then the spin and current response.

\begin{figure}[tb]
\includegraphics[width=0.95\columnwidth]{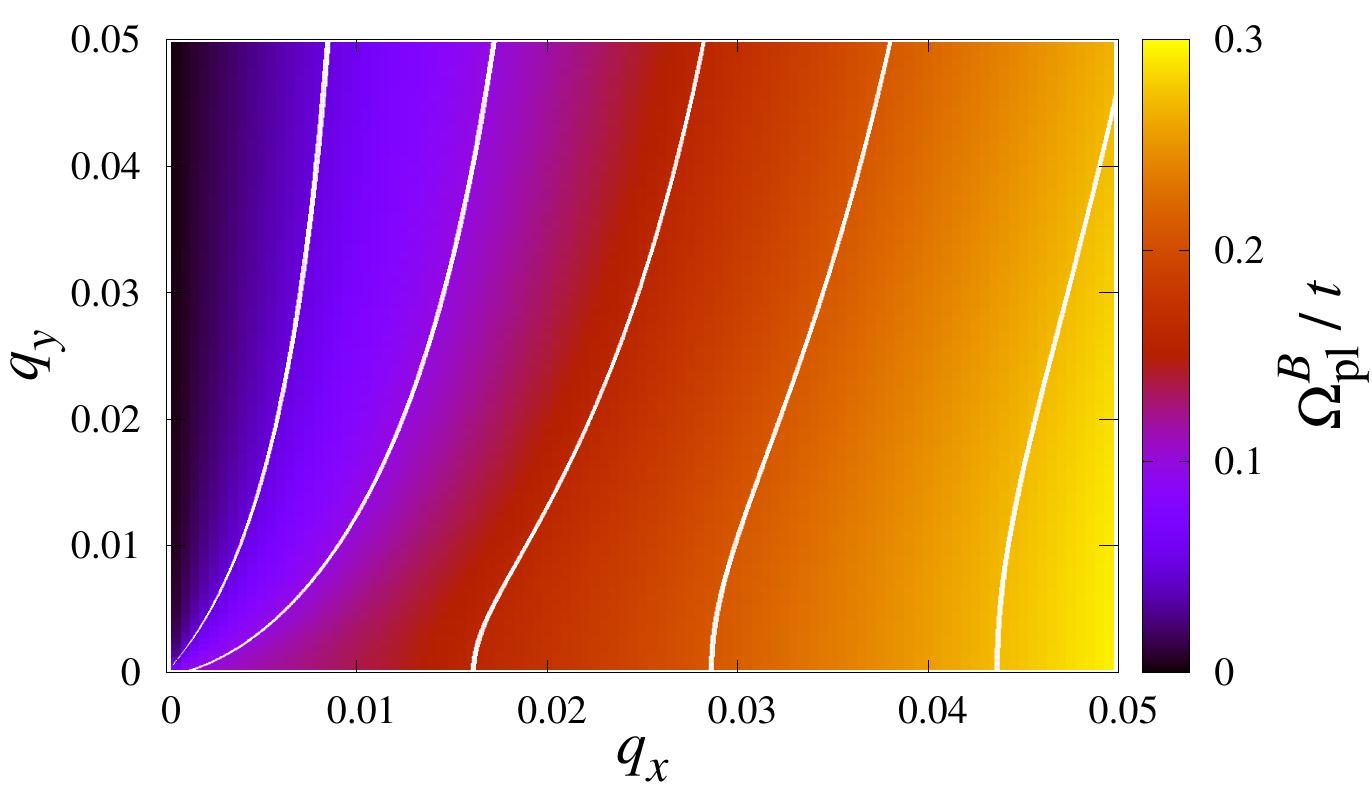}
\caption{Dispersion of the FA plasmons at the bottom surface in the $q_xq_y$-plane. Only one quadrant is shown, the others can be obtained by using $\Omega^B_{\rm pl}(q_x,q_y) = \Omega^B_{\rm pl}(q_x,-q_y) = -\Omega^B_{\rm pl}(-q_x,q_y) = -\Omega^B_{\rm pl}(-q_x,-q_y)$. The effective fine structure constant is taken to be $\alpha_{\kappa}= 0.3$. Constant-frequency contours are shown as white lines. The plasmon group velocity is normal to these lines.}
\label{fig_plasmon} 
\end{figure}

\subsubsection{Density response and Fermi-arc plasmons}
\label{secVB1}

We first consider the interacting density response of the FA states. This allows us to study the effect of FAs on the surface plasmons. Surface plasmons for Weyl semimetals were previously studied in Refs.\ \cite{Song2017,Hofmann2016,Andolina2018,Losic2018,Gorbar2019}. Song and Rudner \cite{Song2017} used classical electrodynamics in a simple phenomenological model to obtain surface plasmons based on the boundary conditions. Hofmann and Das Sarma \cite{Hofmann2016} have studied surface plasmon polaritons using the similar technique deep in the retarded regime, i.e., for surface plasmon wave numbers $q \sim \omega/c$. Gorbar {\it et al.}\ \cite{Gorbar2019} presented a hydrodynamic description of surface collective modes. They found one gapped and one (linear) gapless branch. Lo\v{s}i\'c \cite{Losic2018} calculated the surface-plasmon dispersion in the RPA due to two-dimensional FA states and obtained a $\sqrt{q}$ dispersion. Andolina {\it et al.}\ \cite{Andolina2018} presented a more sophisticated quantum-mechanical calculation of surface-plasmon excitations. The contribution to the surface plasmons from FAs was found to be gapped. However, in all of these works the bulk is assumed to be doped and the surface plasmon is affected by contributions from the bulk states. In the undoped limit, when the chemical potential is at the Weyl nodes, the bulk carrier density vanishes and the surface plasmon is formed only by the FA states. In the following, we investigate this scenario of surface plasmons formed by FA states alone, which we call FA plasmons.

The dispersion of the FA plasmons is given by the zeros of the effective FA dielectric function $\epsilon^{T,B}_{\rm RPA}({\bf q},\omega) = 1 + V_q\, \Pi^{T,B}_{00}({\bf q},\omega)$ \footnote{To be precise, the interacting density response functions depend on the $z$ coordinates and read $\tilde{\Pi}^{T,B}_{00}(z,z',{\bf q}, \omega) = {\Pi}^{T,B}_{00}(z,z',{\bf q}, \omega)/\epsilon^{T,B}_{\rm RPA}(z,z',{\bf q},\omega)$, where the RPA dielectric function is defined as $\epsilon^{T,B}_{\rm RPA}(z,z',{\bf q},\omega) = \delta(z-z') + \frac{2\pi e^2}{\kappa q} \int dz'' \exp(-q|z-z''|)\, {\Pi}^{T,B}_{00}(z'',z',{\bf q}, \omega)$}, where $V_q = 2\pi e^2/\kappa q$ is the Fourier-transformed Coulomb interaction in two dimensions as we are interested in the surface density response. This is valid for ${\bf q}$ small compared to the typical inverse decay length of FA states, which is of the order of $b_y/2\lambda_z$. For consistency, we also expand $\Pi^{T,B}_{00}$ up to the leading term, which yields the FA-plasmon dispersion
\begin{equation}
\Omega^{T,B}_{\rm pl} ({\bf q})
  = \mp \cos \theta_{\bf q} \left( \frac{\alpha_{\kappa} b_y}{\pi} + 2\lambda q \right) , 
\label{plasmon}
\end{equation}
where $\theta_{\bf q} = \arctan(q_y/q_x)$ and the upper (lower) sign pertains to the top (bottom) surface. A detailed quantum-mechanical description of FA plasmons is presented in Appendix \ref{app_plasmon}. The resulting dispersion boils down to Eq.\ (\ref{plasmon}) in the long-wavelength limit. Recall that $\alpha_{\kappa} = e^2/2\kappa \lambda$. The FA-plasmon dispersion for the bottom surface is shown in Fig.\ \ref{fig_plasmon}. The plasmon frequency is an odd function of $q_x$, reflecting the chiral nature of the FA plasmon \cite{Andolina2018}. This can be traced back to the FA dielectric function $\epsilon^{T,B}_{\rm RPA}({\bf q},\omega)$ having only a single zero as a function of frequency $\omega$ for fixed momentum ${\bf q}$. This is different from the two-dimensional electron gas, for which the dielectric function has two zeros at $\pm \omega$.

Evidently, the FA plasmons are highly anisotropic. The plasmon energy is maximal for propagation in the direction parallel to the dispersion of the FA states, i.e., the $x$-direction for our model, and goes to zero in the perpendicular direction. The dispersion has a gap $\alpha_{\kappa} b_y \cos \theta_{\bf q} / \pi$, which is direction dependent and proportional to the separation of the Weyl nodes. The lowest-order nonlocal term is linear in $q$. The group velocity of FA plasmons is also anisotropic and is given by 
\begin{equation}
{\bf v}^{T,B}_{\rm pl} = \hat{\mbox{\boldmath$\theta$}}\, q^{-1}\,
  \frac{\partial\Omega^{T,B}_{\rm pl}}{\partial{\theta}}
  = \pm \frac{\alpha_{\kappa} b_y}{\pi q} \sin \theta_{\bf q}\, \hat{\mbox{\boldmath$\theta$}} ,
\end{equation}
to lowest order. The group-velocity vector is always orthogonal to the constant-frequency contours in Fig.\ \ref{fig_plasmon}. Owing to the opposite sign of the band dispersions for the two surfaces, the chiral FA plasmons move in opposite directions at the two surfaces.

The FA plasmons are distinct from the surface plasmons on a planar surface of a normal metal \cite{Ritchie1957,Ritchie1966,Ritchie1973}. The FA states are much more akin to one-dimensional chiral integer-quantum-Hall edge states in that they are essentially unidirectional and chiral. Indeed, the quantum-Hall edge with long-range Coulomb interaction supports chiral plasmons \cite{Glattli1985,Ashoori1992,Aleiner1994,Hashisaka2018}, like the FAs. 

From an experimental perspective, the FA surface plasmon in WSMs can be investigated by using high-resolution electron energy loss spectroscopy \cite{Chiarello2019} and near-field optical spectroscopy, see Ref.~\cite{Basov2016} for a review. For a typical experimental value of $v_F = 4 \times 10^5\,\mathrm{ms}^{-1}$ \cite{Sushkov2015}, a separation of the Weyl nodes of $0.3\, \mbox{\AA}^{-1}$ \cite{Wang2018}, and assuming the dielectric constant $\kappa$ to be on the order of $10$, the FA-plasmon gap for $\theta_{\bf q} = 0$ turns out to be on the order of $60\,\mathrm{meV}$. This corresponds to a frequency of about $15\,\mathrm{THz}$, in good agreement with experimental results \cite{Chiarello2019}. As the FA plasmons contain information about the bulk band structure, e.g., the separation of Weyl nodes and their Fermi velocities, its experimental study is a promising tool for investigating not only the properties of the FA states but also---so to speak, holographically---of the WSM bulk.

\subsubsection{Spin and current response}

We find that the spin response of the FA states is not influenced by the electron interaction at the RPA level, similarly to the bulk. The physical spin response for the $x$-component is given by $\Pi_2^{T,B}$, whereas the response for the $y$- and $z$-components vanish. As a result, the Pauli spin susceptibility for the FA states also remains unaffected by the interaction. However, the crossed density-spin response is influenced and is now given by
\begin{align}
\widetilde{\Pi}^{T,B,\mu\mu\nu\nu}_{0i}({\bf q}, \omega)
  &= \frac{\Pi^{T,B,\mu\mu\nu\nu}_{0i}({\bf q}, \omega)}{1 + V_q\, \Pi_{00}^{T,B}({\bf q}, \omega)} , \\ \widetilde{\Pi}^{T,B,\mu\mu\nu\nu}_{i0}({\bf q}, \omega)
  &= \frac{\Pi^{T,B,\mu\mu\nu\nu}_{i0}({\bf q}, \omega)}{1 + V_q\, \Pi_{00}^{T,B}({\bf q}, \omega)}.
\end{align}
As discussed previously, the density-spin response gives rise to the magnetoelectric effect. Therefore, the FA magnetoelectric response, which is one of the salient features of the FA states, is enhanced by the inverse FA dielectric function. Consequently, the response will exhibit a pole at the FA plasmon frequency. This behavior is a key observable feature of FA states in interacting WSMs.

To give a specific example, let us consider an inhomogeneous oscillating electric field, defined by
\begin{equation}
{\bf E} = E_0\, e^{i (q_x x - \omega t)}\, \hat{\bf y} ,
\end{equation}
applied to the WSM surface. Owing to the magnetoelectric effect, the electric field induces a magnetization of the FAs, which is given by
\begin{align}
{\bf M}^{T,B}(q_x,\omega) &= \mp \frac{e \mu_B }{24 \pi^2}\, (g^\oplus - g^\ominus) \nonumber \\
&{}\times \frac{1}{\epsilon^{T,B}_{\rm RPA}(q_x,\omega)}\,
  \frac{q_x}{2\lambda q_x \pm (\omega + i\delta)}\, E_0\, \hat{\bf z}.
\end{align}
Therefore, the magnetization will exhibit two peaks in the $(q_x,\omega)$ plane, at the plasmon frequency $\Omega_{\rm pl}^{T,B}$ and at $\mp 2 \lambda q_x$. This is a striking feature of FA states and provides a unique tool to probe the surface-plasmon dispersion.

Finally, the current-current correlation function is found to be similarly enhanced by interactions,
\begin{equation}
\widetilde{\Pi}^{T,B}_{j_x j_x}({\bf q}, \omega)
  = \frac{\Pi^{T,B}_{j_x j_x}({\bf q}, \omega)}{1 + V_q\, \Pi_{00}^{T,B}({\bf q}, \omega)}.
\end{equation}
Evidently, the proportionality between current-current correlation and density susceptibility holds true for interacting electrons as well.

\section{Summary and conclusions}
\label{sec6}

To summarize, we have investigated the dynamical density and spin response of FA states of time-reversal-symmetry-breaking WSMs. We have obtained a model for a WSM by closing the band gap in a topological insulating state, leading to a Dirac semimetal, and breaking time-reversal symmetry explicitly. This approach is convenient for analytical evaluations but the general conclusions are not expected to depend on the specific model since they rely on the topological invariants of Weyl nodes and the universal linear low-energy dispersion of FA surface states. We have obtained the evanescent wave functions of the FA states and analytical expressions for all components of the wave-vector- and frequency-dependent composite density-spin response tensor. The penetration of the FA states into the bulk, which becomes large for momenta close to the Weyl nodes, has been found to be crucial for the correct low-frequency behavior.

We have then examined observable consequences of our results for the electric, magnetic, and coupled magnetoelectric susceptibilities as well as for the optical conductivity and the anomalous Hall effect. In particular, we have found that the FAs exhibit a chiral magnetoelectric effect. Also, the FA states lead to an anomalous Hall effect. For \emph{time-reversal-symmetric} WSMs with broken inversion symmetry, FAs at a given surface come in pairs related by time reversal. The resulting linear response is the sum of their contributions. Consequently, the electric and magnetic susceptibilities, which are even under time reversal, add up, whereas the mixed magnetoelectric susceptibility, which is odd under time reversal, vanishes.

Based on the full response tensor, we have studied the impact of the electron-electron Coulomb interaction on the FA and bulk response, within the RPA. While the RPA spin response is unaffected by the interaction effects, the density and current susceptibilities are strongly renormalized. The spectrum of surface density excitations for FA states contains chiral FA plasmons, whose dispersion is highly anisotropic and yields information about the electronic structure of WSMs. Moreover, the FA magnetoelectric effect is also renormalized by the FA plasmon dispersion and will show resonance-like behavior when frequency and momentum match the FA-plasmon dispersion. The FA plasmons of time-reversal-symmetric WSMs are similar to our case, except for the vanishing magnetoelectric effect. The reason for this is that only the charge susceptibility appears in the denominators in the RPA expressions, for example in Eq.\ (\ref{RPA}).

We hope that our study of dynamical response functions will motivate experimental studies of the spin response of FA states, similar to experiments on bulk WSMs in Ref.\ \cite{Ma2019}. Our work should also be useful for exploring nonlocal transport and optical properties of the FA surface states. Moreover, the magnetoelectric effect, the anisotropic FA plasmon, and their interplay have the potential to lead to smoking-gun experimental evidence for the FA states and thus for the presence of Weyl nodes in the bulk.

\acknowledgments

The authors thank G.\,M. Andolina, M. Polini, and K. Roychowdhury for useful discussions. Financial support by the Deut\-sche For\-schungs\-gemein\-schaft through project A04 of Collaborative Research Center SFB 1143 and through the Cluster of Excellence on Complexity and Topology in Quantum Matter ct.qmat (EXC 2147) is gratefully acknowledged.

\appendix

\section{Eigenstates near Weyl nodes}
\label{app_bulk}

Here, we describe the eigenstates near the Weyl nodes, based on the Hamiltonian in Eq.\ (\ref{hamil2}). In the vicinity of the node at $(0, b_y/2\lambda,0)$, the dispersion is given by
\begin{equation}
E_{\pm}({\bf k})
  = \pm \varepsilon({\bf k})
  \equiv \pm \sqrt{4\lambda^2 \tilde{k}_x^2 + 4\lambda^2 \tilde{k}_y^2 + 4\lambda_z^2 \tilde{k}_z^2} ,
\label{dispersion_bulk_1}
\end{equation}
where $(\tilde{k}_x,\tilde{k}_y,\tilde{k}_z)$ is the momentum relative to the node. We write $2 \lambda \tilde{k}_x = \varepsilon({\bf k}) \sin \theta_k \cos \phi_k$, $2 \lambda \tilde{k}_y = \varepsilon({\bf k}) \sin \theta_k \sin \phi_k$, and $2 \lambda_z \tilde{k}_z = \varepsilon({\bf k}) \cos \theta_k$ such that the polar angles $\theta_k$ and $\phi_k$ parametrize constant-energy surfaces. With these substitutions, the periodic parts of the Bloch states are given by
\begin{align}
|\phi^{N}_{+}({\bf k})\rangle = \left( \begin{array}{c}
  \sin\frac{\theta_\mathbf{k}}{2}\, e^{-i \varphi_\mathbf{k}/2} \\[0.7ex]
  -\sin\frac{\theta_\mathbf{k}}{2}\, e^{-i \varphi_\mathbf{k}/2} \\[0.7ex]
  -\cos\frac{\theta_\mathbf{k}}{2}\, e^{i \varphi_\mathbf{k}/2} \\[0.7ex]
  \cos\frac{\theta_\mathbf{k}}{2}\, e^{i \varphi_\mathbf{k}/2} 
  \end{array} \right) , \\
|\phi^{N}_{-}({\bf k})\rangle = \left( \begin{array}{c}
  -\cos\frac{\theta_\mathbf{k}}{2}\, e^{-i \varphi_\mathbf{k}/2} \\[0.7ex]
  \cos\frac{\theta_\mathbf{k}}{2}\, e^{-i \varphi_\mathbf{k}/2} \\[0.7ex]
  - \sin\frac{\theta_\mathbf{k}}{2}\, e^{i \varphi_\mathbf{k}/2}\\[0.7ex]
  \sin\frac{\theta_\mathbf{k}}{2}\, e^{i \varphi_\mathbf{k}/2}\\[0.7ex]
  \end{array}\right).
\end{align}
Here, the superscript $N$ refers to the fact that the node at $(0,b_y/2\lambda,0)$ has negative chirality.
Similarly, the periodic parts of the eigenvectors near the positive-chirality node at $(0,-b_y/2\lambda,0)$ is given by
\begin{align}
|\phi^{P}_{+}({\bf k})\rangle = \left( \begin{array}{c}
  \cos\frac{\theta_\mathbf{k}}{2}\, e^{-i \varphi_\mathbf{k}/2} \\[0.7ex]
  \cos\frac{\theta_\mathbf{k}}{2}\, e^{-i \varphi_\mathbf{k}/2} \\[0.7ex]
  \sin\frac{\theta_\mathbf{k}}{2}\, e^{i \varphi_\mathbf{k}/2} \\[0.7ex]
  \sin\frac{\theta_\mathbf{k}}{2}\, e^{i \varphi_\mathbf{k}/2}
  \end{array} \right) , \\
|\phi^{P}_{-}({\bf k})\rangle = \left( \begin{array}{c}
  \sin\frac{\theta_\mathbf{k}}{2}\, e^{-i \varphi_\mathbf{k}/2} \\[0.7ex]
\sin\frac{\theta_\mathbf{k}}{2}\, e^{-i \varphi_\mathbf{k}/2} \\[0.7ex]
  - \cos\frac{\theta_\mathbf{k}}{2}\, e^{i \varphi_\mathbf{k}/2}\\[0.7ex]
  - \cos\frac{\theta_\mathbf{k}}{2}\, e^{i \varphi_\mathbf{k}/2}
  \end{array}\right) .
\end{align}
Note that the bulk states are eigenspinors of $\sigma_0 \otimes \tau_x$.

\section{Calculation of Fermi-arc states}
\label{app_FA}

In this appendix, we present the derivation of the wave functions of FA states. We start by making the following assumptions.

(i) The top and bottom surfaces can be solved independently. This means that the WSM slab is assumed to be sufficiently thick so that the states at the two surfaces are decoupled.

(ii) The vacuum is an ordinary insulator with a very large mass so that the vacuum Hamiltonian is given by Eq.\ (\ref{hamil2}) with $\epsilon \gg t, \lambda, \lambda_z$. The top surface separates the WSM at $z<0$ from the vacuum at $z>0$ and is characterized by a $z$-dependent mass $M(z) = (6t - 2t \sum_\alpha \cos k_\alpha)\, \Theta(-z) + \mathcal{M}\, \Theta(z)$, where $\Theta(z)$ is the Heaviside step function and $\mathcal{M}$ is the large vacuum mass. Similarly, the mass for the bottom surface is $M(z) = (6t - 2t \sum_\alpha \cos k_\alpha)\, \Theta(z) + \mathcal{M}\, \Theta(-z)$.

(iii) The FA wave functions are bound to the surfaces and consists of a periodic term in ${\bf r}_\parallel = (x,y)$ and a evanescent factor in $z$,
\begin{equation}
\Phi^{T,B}({\bf k}_\parallel, {\bf r})
  = \phi^{T,B}({\bf k}_\parallel,z)\, e^{i{\bf k}_\parallel \cdot {\bf r}_\parallel} ,
\end{equation}
with $T$, $B$ referring to the top and bottom surfaces, respectively, and ${\bf k}_\parallel = (k_x,k_y)$. 

First, we calculate the FA states for the top surface. The FA states exist for $-b_y/2\lambda \leq k_y \leq b_y / 2\lambda$ and have the dispersion $E^{T}({\bf k}) = - 2\lambda k_x$ \cite{Note1}. Linearizing the Hamiltonian in Eq.\ (\ref{hamil2}) and replacing $k_z$ by $-i\,\partial/\partial z \equiv -i\partial_z$, the Dirac-type equation for the evanescent part within the WSM reads
\begin{widetext}
\begin{equation}
\left(\begin{array}{c c c c}
0 & - 2i \lambda_z \partial_z & -ib_y & 2\lambda (k_x - i k_y) \\
 - 2i \lambda_z \partial_z & 0 & 2\lambda (k_x - i k_y) & -ib_y \\
ib_y & 2\lambda (k_x + i k_y) & 0 & 2i \lambda_z \partial_z \\
2\lambda (k_x + i k_y) & ib_y & 2i \lambda_z \partial_z & 0 
\end{array}\right)
\left(\begin{array}{c}
\phi^{T}_1  \\
\phi^{T}_2 \\
\phi^{T}_3\\
\phi^{T}_4
\end{array}\right) = - 2\lambda k_x \left(\begin{array}{c}
\phi^{T}_1  \\
\phi^{T}_2 \\
\phi^{T}_3\\
\phi^{T}_4
\end{array}\right).
\label{dirac_eq1}
\end{equation}
Using the ansatz $\phi^T_i = \psi^T_i e^{\kappa z}$, we obtain a linear homogeneous system of equations for the $\psi^T_i$ given by
\begin{equation}
\left(\begin{array}{c c c c}
2\lambda k_x & - 2i \lambda_z \kappa & -ib_y & 2\lambda (k_x - i k_y) \\
 - 2i \lambda_z \kappa & 2\lambda k_x & 2\lambda (k_x - i k_y) & -ib_y \\
ib_y & 2\lambda (k_x + i k_y) & 2\lambda k_x & 2i \lambda_z \kappa \\
2\lambda (k_x + i k_y) & ib_y & 2i \lambda_z \kappa & 2\lambda k_x 
\end{array}\right)
\left(\begin{array}{c}
\psi^{T}_1  \\
\psi^{T}_2 \\
\psi^{T}_3\\
\psi^{T}_4
\end{array}\right) = 0.
\label{dirac_eq2}
\end{equation}
\end{widetext}
For a nontrivial solution to exit, the determinant of the coefficient matrix must vanish, which yields four solutions,
\begin{equation}
\kappa \in \left\{ \frac{b_y - 2\lambda k_y}{2\lambda_z},
  \frac{b_y + 2\lambda k_y}{2\lambda_z},
  -\frac{b_y - 2\lambda k_y}{2\lambda_z},
  -\frac{b_y + 2\lambda k_y}{2\lambda_z} \right\} .
\end{equation}
Only the first two values are relevant for the top surface since the wave function should vanish for $z \to -\infty$. Then the rank of the $4\times 4$ matrix turns out to be three so that there is only one linearly independent solution of Eq.\ (\ref{dirac_eq2}) for each value of $\kappa$. The resultant evanescent spinor is the linear combination of the two and is given by
\begin{equation}
\phi^{T} ({\bf k}_\parallel,z) = \alpha \left(\begin{array}{c}
1 \\ -1 \\ 1 \\ -1
\end{array}\right) e^{\frac{b_y - 2\lambda k_y}{2\lambda_z}z} + \beta \left(\begin{array}{c}
1 \\ 1 \\ -1 \\ -1
\end{array}\right) e^{\frac{b_y + 2\lambda k_y}{2\lambda_z}z}.
\end{equation} 
To calculate the coefficients $\alpha$ and $\beta$, we consider the wave function at the vacuum side. Using that $\mathcal{M} \gg t,\lambda,\lambda_z$, an analogous derivation yields the evanescent spinor in the vacuum,
\begin{equation}
\phi^{T}_{\rm vac}({\bf k}_\parallel,z) = A \left(\begin{array}{c}
-1 \\ - i \\ i \\ 1
\end{array}\right) e^{-\frac{\mathcal{M}}{2\lambda_z}z} + B \left(\begin{array}{c}
1 \\ i \\ i \\ 1
\end{array}\right) e^{-\frac{\mathcal{M}}{2\lambda_z}z} .
\end{equation} 
Using continuity at $z = 0$, we get $B = 0$, $\alpha = -(1-i)A/2$, and $\beta = - (1+i)A/2$. Taking $A = -1$, the evanescent part of the FA wave function for the top surface, within the WSM, is given by
\begin{align}
\phi^{T}({\bf k}_\parallel,z)
  &= \sqrt{\frac{b_y^2 - 4 \lambda^2 k_y^2}{4 b_y \lambda_z}}\:
   \left[ \frac{1 - i}{2} \left(\begin{array}{c}
     1 \\ -1 \\ 1 \\ -1
   \end{array}\right) e^{\frac{b_y - 2\lambda k_y}{2\lambda_z}z} \right. \nonumber \\
&{}+ \left. \frac{1 + i}{2} \left(\begin{array}{c}
     1 \\ 1 \\ -1 \\ -1
   \end{array}\right) e^{\frac{b_y + 2\lambda k_y}{2\lambda_z}z} \right] .
\end{align}
Following the same procedure and noting that only negative values of $\kappa$ are relevant, the evanescent part of the wave function for the bottom surface, within the WSM, is found to be
\begin{align}
\phi^{B}({\bf k}_\parallel,z)
  &= \sqrt{\frac{b_y^2 - 4 \lambda^2 k_y^2}{4 b_y \lambda_z}}\:
    \left[ \frac{1+i}{2} \left(\begin{array}{c}
      1 \\ -1 \\ -1 \\ 1
    \end{array}\right) e^{-\frac{b_y - 2\lambda k_y}{2\lambda_z}z} \right. \nonumber \\
&{}+ \left. \frac{1-i}{2} \left(\begin{array}{c}
      1 \\ 1 \\ 1 \\ 1
    \end{array}\right) e^{-\frac{b_y + 2\lambda k_y}{2\lambda_z}z} \right] ,
\end{align} 
with dispersion $E^{B}({\bf k}) = 2 \lambda k_x$.

\section{Fermi-arc states for generic Hamiltonian}
\label{app_generalization}

In the following, we argue that any WSM model with two-valued orbital degrees of freedom will have qualitatively similar FA states as in Eqs.\ (\ref{wave_function_FA.1}) and (\ref{wave_function_FA.2}). We consider the following generic Hamiltonian for a WSM:
\begin{equation}
\mathcal{H}_0 = [ \mathcal{F}_1({\bf k})\, \sigma_x + \mathcal{F}_2({\bf k})\, \sigma_y
  + \mathcal{F}_3({\bf k})\, \sigma_z ] \otimes \tau_x .
\end{equation}
$\mathcal{P}$ and $\mathcal{T}$ symmetry require that $\mathcal{F}_1$ is odd in $k_x$ and even in $k_y$ and $k_z$,  $\mathcal{F}_2$ is odd in $k_y$ and even in $k_z$ and $k_x$, and $\mathcal{F}_3$ is odd in $k_z$ and even in $k_x$ and $k_y$. The $\mathcal{T}$ symmetry is broken by the additional term $b_y \sigma_y \otimes \tau_0$.

To calculate surface states in the $k_xk_y$-plane, we carry out the following expansions to lowest order in $k_z$ and replace $k_z \to - i \partial_z$:
\begin{align}
\mathcal{F}_1 &= f_1(k_x,k_y) , \\
\mathcal{F}_2 &= f_2(k_x,k_y) , \\
\mathcal{F}_3 &= - i \, f_3(k_x,k_y) \, \partial_z. 
\end{align}
The resulting Hamiltonian reads 
\begin{align}
\mathcal{H} = 
\left(\begin{array}{c c c c}
0 & - i f_3 \partial_z & -ib_y & f_1 - i f_2 \\
- i f_3 \partial_z & 0 & f_1 - i f_2 & -ib_y \\
ib_y & f_1 + i f_2 & 0 & i f_3 \partial_z \\
f_1 + i f_2 & ib_y & i f_3 \partial_z & 0 
\end{array}\right).
\end{align}
The locations of the Weyl nodes are given by $b_y \pm |f_2 (k_x,k_y)| = 0$ and the dispersion for the top and bottom surfaces are $E^{T,B}({\bf k}_\parallel)=\mp |f_1 (k_x,k_y)| $.

With a derivation similar to Appendix \ref{app_FA}, and using an ansatz of the from $\phi^{T,B}_i = \psi^{T,B}_i e^{\kappa z}$ for evanescent modes, we find
\begin{align}
\kappa \in \left\{ \frac{|f_3|}{b_y -|f_2|}, \frac{|f_3|}{b_y +|f_2|},
  -\frac{|f_3|}{b_y -|f_2|}, -\frac{|f_3|}{b_y +|f_2|} \right\} .
\end{align}  
Defining $\kappa_1 = |f_3|/(b_y -|f_2|)$ and $\kappa_2 = |f_3|/(b_y +|f_2|)$, the evanescent part of the normalized wave functions for the top and bottom surfaces are given by
\begin{widetext}
\begin{align}
\phi^{T} ({\bf k}_\parallel,z) &= \sqrt{\frac{\kappa_1 \kappa_2}{\kappa_1 + \kappa_2}}\:
  \left[ \frac{1 - i}{2} \left(\begin{array}{c}
    1 \\ -1 \\ 1 \\ -1
  \end{array}\right) e^{\kappa_1 z} + \frac{1 + i}{2} \left(\begin{array}{c}
    1 \\ 1 \\ -1 \\ -1
  \end{array}\right) e^{\kappa_2 z} \right] , \\
\phi^{B} ({\bf k}_\parallel,z) &= \sqrt{\frac{\kappa_1 \kappa_2}{\kappa_1 + \kappa_2}}\:
  \left[ \frac{1 + i}{2} \left(\begin{array}{c}
    1 \\ -1 \\ -1 \\ 1
  \end{array}\right) e^{-\kappa_1 z} + \frac{1 - i}{2} \left(\begin{array}{c}
    1 \\ 1 \\ 1 \\ 1
  \end{array}\right) e^{-\kappa_2 z} \right],
\end{align}
\end{widetext}
where $\kappa_1$ and $\kappa_2$ are functions of $k_x$ and $k_y$. Evidently, the FA states are eigenstates of $\sigma_x \otimes \tau_x$.

Alternatively, $\mathcal{T}$ symmetry can be broken by adding the term $b_x \sigma_x \otimes \tau_0$ to the Hamiltonian $\mathcal{H}_0$. In this case, the FA states become eigenstates of $\sigma_y \otimes \tau_x$ with the same decay length scales $1/\kappa_1$ and $1/\kappa_2$.

\section{Dynamical response of bulk states}
\label{app_bulk_response}

Here, we briefly discuss the dynamical response functions for the bulk states. For chemical potential close to the energy of the Weyl nodes, only the states near the two nodes contribute to the response functions. Since the low-energy physics is governed by the two bands constituting the Weyl cones, we calculate the response for states in the vicinity of the nodes up to a cutoff energy $\varepsilon_c$. The technique is similar to Ref.\ \cite{Ghosh2019} for a ${\bf k}\cdot{\bf p}$ Hamiltonian and we therefore do not present it in detail. 

\subsection{Density response}

The density response $\Pi_{00}$ is determined by the density-density correlations and we have $\Pi_{00}^{\oplus \oplus \oplus \oplus} = \Pi_{00}^{\oplus \oplus \ominus \ominus} = \Pi_{00}^{\ominus \ominus \oplus \oplus} = \Pi_{00}^{\ominus \ominus \ominus \ominus} = \Pi_{00}^{\oplus \ominus \oplus \ominus} = \Pi_{00}^{\ominus \oplus \ominus \oplus} = \Pi_{00}^{\oplus \ominus \ominus \oplus} = \Pi_{00}^{\ominus \oplus \oplus \ominus} \equiv \Pi_{00}^{\rm bulk}$ and the other terms are zero. $\Pi_{00}^{\rm bulk}$ can be written as the sum of a contribution from the undoped system (``intrinsic'') and a contribution due to doping (``extrinsic'') as $\Pi_{00}^{\rm bulk}(\mathbf{q},\omega) = \Pi_{00}^{\rm in,bulk}(\mathbf{q},\omega) + \Pi_{00}^{\rm ex,bulk}(\mathbf{q},\omega)$. The imaginary and real parts of the intrinsic contribution are given by
\begin{align}
{\rm Im}\,\Pi_{00}^{\rm in,bulk}(\mathbf{q},\omega) &= \frac{({\bf v}_F\cdot{\bf q})^2}{12 \pi v_F^3}\, \Theta(\omega - {\bf v}_F\cdot{\bf q}) ,
\label{IPi00in} \\
{\rm Re}\,\Pi_{00}^{\rm in,bulk}(\mathbf{q},\omega) &= \frac{({\bf v}_F\cdot{\bf q})^2}{12 \pi^2 v_F^3}\,   \ln\left| \frac{4 \varepsilon_c^2}{({\bf v}_F\cdot{\bf q})^2 - \omega^2} \right| ,
\label{RPi00in}
\end{align}
respectively. For electron doping, $\mu > 0$, the imaginary and real parts of the extrinsic contribution read 
\begin{widetext}
\begin{align}
&{\rm Im}\,\Pi^{\rm ex,bulk}_{00}(\mathbf{q},\omega)
  = \frac{1}{4 \pi v_F}\, \bigg[ \Theta({\bf v}_F\cdot{\bf q} - \omega)
    \Big( [\alpha({\bf q},\omega) - \alpha({\bf q},-\omega)] \Theta(2\mu - {\bf v}_F\cdot{\bf q} - \omega)
    \nonumber \\
&\quad{} + \alpha({\bf q},\omega) \Theta(2\mu - {\bf v}_F\cdot{\bf q} + \omega)
    \Theta({\bf v}_F\cdot{\bf q} + \omega - 2\mu) \Big) \nonumber \\
&\quad{}+ \Theta(\omega - {\bf v}_F\cdot{\bf q})
  \bigg( {-} \alpha(-{\bf q},-\omega) \Theta(2\mu + {\bf v}_F\cdot{\bf q} - \omega)
  \Theta({\bf v}_F\cdot{\bf q} + \omega - 2\mu)
  - \frac{({\bf v}_F\cdot{\bf q})^2}{3v_F^2}\, \Theta(2\mu - {\bf v}_F\cdot{\bf q} - \omega) \bigg) \bigg] , \\ 
&{\rm Re}\,\Pi^{\rm ex,bulk}_{00}(\mathbf{q},\omega)
  = \frac{1}{4\pi^2 v_F} \bigg[ \frac{8\mu^2}{3v_F^2} - \alpha({\bf q},\omega)\beta({\bf q},\omega)
  - \alpha(-{\bf q},\omega)\beta(-{\bf q},\omega) - \alpha({\bf q},-\omega)\beta({\bf q},-\omega)
  - \alpha(-{\bf q},-\omega)\beta(-{\bf q},-\omega) \bigg] ,
\label{RPirhorhoex}
\end{align}
where
\begin{align}
\alpha({\bf q},\omega) \equiv \frac{1}{12v_F^2 ({\bf v}_F\cdot{\bf q})} \left[(2\mu+\omega)^3  -3({\bf v}_F\cdot{\bf q})^2 (2\mu+\omega) + 2 ({\bf v}_F\cdot{\bf q})^3 \right]
\label{def_alpha}
\end{align}
\end{widetext}
and
\begin{equation}
\beta({\bf q},\omega) \equiv \ln \left| \frac{2\mu + \omega - ({\bf v}_F\cdot{\bf q})}{({\bf v}_F\cdot{\bf q}) - \omega} \right| ,
\label{def_beta}
\end{equation}
with ${\bf v}_F = (2\lambda,2\lambda,2\lambda_z)$ being the Fermi-velocity vector near the Weyl nodes.

\subsection{Spin response}

Similarly, for the spin response, terms of the form $\Pi_{ij}^{\mu \mu \nu \nu}$, $\Pi_{ij}^{\mu \bar{\mu} \mu \bar{\mu}}$, and $\Pi_{ij}^{\mu \bar{\mu} \bar{\mu} \mu}$ are equal to $\Pi_{ij}^{\rm bulk}$ and the other terms vanish. The spin response consists of diagonal terms ($i=j$) and off-diagonal terms ($i \ne j$). The diagonal components can be further decomposed into longitudinal ($\Pi_{ll}$ for ${\bf q} = q\, \hat{\bf l}$) and transverse ($\Pi_{mm}$ and $\Pi_{nn}$ for ${\bf q} = q\, \hat{\bf l}$) response, where $\hat{\bf l}$, $\hat{\bf m}$, and $\hat{\bf n}$ are three orthogonal coordinate axes forming a right-handed system. The longitudinal components can be written in terms of the density response as
\begin{equation}
\Pi_{ll}^{\rm bulk}(q\, \hat{\mathbf{l}},\omega) = \frac{\omega^2}{({\bf v}_F \cdot {\bf q})^2}\,
  \Pi_{00}^{\rm bulk}(q\, \hat{\mathbf{l}},\omega) .
\label{Pi_zz}
\end{equation}
On the other hand, for the transverse response, the proportionality to the density response only holds for the intrinsic part,
\begin{equation}
\Pi_{mm}^{\rm in,bulk}(q\, \hat{\mathbf{l}},\omega) = \frac{\omega^2 - ({\bf v}_F \cdot {\bf q})^2}
  {({\bf v}_F \cdot {\bf q})^2}\, \Pi_{00}^{\rm in,bulk}(q\, \hat{\mathbf{l}},\omega),
\end{equation}
while the extrinsic part has a more complicated form,
\begin{widetext}
\begin{align}
&\mathrm{Im}\,\Pi_{mm}^{\rm ex,bulk}(q\, \hat{\mathbf{l}},\omega)
  = \frac{\omega^2 - ({\bf v}_F\cdot {\bf q})^2}{16 \pi ({\bf v}_F\cdot {\bf q})^3}\,
    \bigg[ \Theta({\bf v}_F\cdot {\bf q} - \omega)
    \Big( [\gamma({\bf q},\omega) - \gamma({\bf q},-\omega)] \Theta(2\mu - {\bf v}_F\cdot {\bf q} - \omega)
    \nonumber \\
&\quad{}+ \gamma({\bf q},\omega) \Theta(2\mu - {\bf v}_F\cdot {\bf q} + \omega) \Theta({\bf v}_F\cdot {\bf q} + \omega - 2\mu) \Big) \nonumber \\
&\quad{}+ \Theta(\omega - {\bf v}_F\cdot {\bf q}) \bigg( \gamma(-{\bf q},-\omega)
  \Theta(2\mu + {\bf v}_F\cdot {\bf q} - \omega)\Theta({\bf v}_F\cdot {\bf q} + \omega - 2\mu)
  + \frac{4q^2}{3}\, \Theta(2\mu - {\bf v}_F\cdot {\bf q} - \omega) \bigg) \bigg] , \\
&\mathrm{Re}\,\Pi_{mm}^{\rm ex,bulk}(q\, \hat{\mathbf{l}},\omega)
  = - \frac{\omega^2 - ({\bf v}_F\cdot {\bf q})^2}{2({\bf v}_F\cdot {\bf q})^2}\,
  \Pi_{00}^{\rm ex,bulk}(q\, \hat{\mathbf{l}},\omega)
  - \frac{\mu^2}{2 \pi^2 v_F^3} \nonumber \\
&\quad{} - \frac{\omega^2 - ({\bf v}_F\cdot {\bf q})^2}{16\pi^2 v_F^3 ({\bf v}_F\cdot {\bf q})}\,
  \Bigg( \Theta({\bf v}_F\cdot {\bf q}-\mu)
    \big[ \xi({\bf q},\omega) \beta(-{\bf q},\omega) + \xi({\bf q},-\omega) \beta(-{\bf q},-\omega)
    - \xi(-{\bf q},\omega) \beta({\bf q},\omega) - \xi(-{\bf q},-\omega) \beta({\bf q},-\omega) \big] \nonumber \\
&\quad{} + \Theta(\mu-{\bf v}_F\cdot {\bf q}) \bigg[ (2\mu+\omega)
  \ln \left| \frac{\xi({\bf q},\omega)}{\xi(-{\bf q},\omega)} \right|
    + (2\mu-\omega) \ln \left| \frac{\xi({\bf q},-\omega)}{\xi(-{\bf q},-\omega)} \right|
    - 2 \omega \ln \left| \frac{{\bf v}_F\cdot {\bf q} + \omega}{{\bf v}_F\cdot {\bf q} - \omega} \right|
     \nonumber \\
&\quad{}+ ({\bf v}_F\cdot {\bf q}) \big[ \zeta({\bf q},\omega) + \zeta(-{\bf q},\omega)
      + \zeta({\bf q},-\omega) + \zeta(-{\bf q},-\omega) \big] \bigg] \Bigg) ,
\end{align}
where $\gamma({\bf q},\omega) \equiv 2q\, \alpha({\bf q},\omega) + q^2\, (2 \mu - {\bf v}_F\cdot {\bf q} + \omega)$, $\xi({\bf q},\omega) \equiv 2\mu + {\bf v}_F\cdot {\bf q} + \omega$, and
\begin{equation}
\zeta({\bf q},\omega)
  \equiv \ln\left| \frac{2\mu + {\bf v}_F\cdot {\bf q} + \omega}{{\bf v}_F\cdot {\bf q} + \omega} \right| .
\end{equation}
\end{widetext}
On the other hand, the off-diagonal components of  the spin response tensor can be written in terms of the diagonal components as
\begin{equation}
\Pi_{lm}({\bf q},\omega) = \left[ \Pi_{ll}(q\, \hat{\bf l},\omega)
  - \Pi_{ll}(q\, \hat{\bf m},\omega) \right] \frac{q_l q_m}{q^2}.
\end{equation}

\subsection{Coupled density-spin response}

Due to the coupling between spin and momentum of Weyl fermions, the density and spin degrees of freedom are strongly coupled and the crossed density-spin response is large. However, the same effect causes the intra-orbital and inter-orbital contributions to the density-spin response to change sign with chirality and therefore they do not physically manifest for our model in equilibrium. When the WSM is driven out of equilibrium by the application of non-orthogonal electric and magnetic fields, there will be a non-zero density-spin response, as discussed in Ref.\ \cite{Ghosh2019}. The non-equilibrium case is beyond the scope of this work. However, it is important to note that terms such as $\Pi_{0l}^{\mu\mu\mu\bar{\mu}}(q\, \hat{\bf l},\omega)$ survive even in equilibrium and assume the value $\omega/({\bf v}_F\cdot {\bf q})\, \Pi_{00}^{\rm bulk}$. These terms are irrelevant at the non-interacting level because they do not enter the crossed density-spin response functions in Eqs.\ (\ref{eq_sus_electromagnetic}) and (\ref{eq_sus_magnetoelectric}) but they do affect the current response for interacting electrons, as we discuss in the main part.

\section{Calculation of density response}
\label{app_Pi00}

In this appendix, we discuss the calculation of the density response tensor from the density-density correlation function using Eq.\ (\ref{response_FA}) with the FA wave functions given in Eqs.\ (\ref{wave_function_FA.1}) and (\ref{wave_function_FA.2}). We find that $\Pi_{00}^{\oplus \oplus \oplus \oplus} = \Pi_{00}^{\oplus \oplus \ominus \ominus} = \Pi_{00}^{\ominus \ominus \oplus \oplus} = \Pi_{00}^{\ominus \ominus \ominus \ominus} \equiv \Pi^{B}_1$ is given by
\begin{widetext}
\begin{align}
\Pi^{B}_1 &= \frac{1}{(2\pi)^2} \frac{q_x}{2\lambda q_x - \omega - i \delta} \int_{-{b_y}/{2\lambda}}^{{b_y}/{2\lambda} - q_y} dk_y\,
  \frac{b_y^2 - 4 \lambda^2 k_y^2}{4 b_y \lambda_z}\,
  \frac{b_y^2 - 4 \lambda^2 (k_y + q_y)^2}{4 b_y \lambda_z} \nonumber \\
&\quad{}\times \int_0^{\infty} dz\, e^{-(b_y/\lambda_z)\,z}\, 2\cosh \Big(\frac{\lambda}{\lambda_z} (2k_y + q_y)
  z\Big)
  \int_0^{\infty} dz'\, e^{-(b_y/\lambda_z)\,z'}\, 2\cosh \Big(\frac{\lambda}{\lambda_z} (2k_y + q_y) z'\Big)
  \nonumber \\
&= \frac{1}{(2\pi)^2} \frac{q_x}{2\lambda q_x - \omega - i \delta} \, \frac{1}{8}
  \left[ \frac{2b_y}{\lambda} - 4q_y + \frac{\lambda q_y^2}{b_y}
  + \frac{\lambda^2 q_y^3}{b_y^2} + \frac{\lambda^3 q_y^4}{b_y^3}\,
    \arctanh \left(1-\frac{q_y \lambda}{b_y}\right) \right]
\end{align}
for the bottom surface, assuming $q_x,q_y > 0$. For the top surface, the corresponding results are obtained by replacing $\omega \to - \omega$. Similar calculations for terms such as $\Pi^{\mu\bar{\mu}\mu\mu}$ reveal that the response function vanish when one of the four orbital indices is different from the other three. On the other hand, we find that $\Pi_{00}^{\oplus\ominus\oplus\ominus} = \Pi_{00}^{\oplus\ominus\ominus\oplus} = \Pi_{00}^{\ominus\oplus\ominus\oplus} = \Pi_{00}^{\ominus \oplus \oplus \ominus} \equiv \Pi^{B}_2$ are given by
\begin{align}
\Pi^{B}_2 &= \frac{1}{(2\pi)^2} \frac{q_x}{2\lambda q_x - \omega - i \delta} \int_{-{b_y}/{2\lambda}}^{{b_y}/{2\lambda} - q_y} dk_y\,
  \frac{b_y^2 - 4 \lambda^2 k_y^2}{4 b_y \lambda_z}\,
  \frac{b_y^2 - 4 \lambda^2 (k_y + q_y)^2}{4 b_y \lambda_z} \nonumber \\
&\quad{}\times \int_0^{\infty} dz\, e^{-(b_y/\lambda_z)\,z}\, 2\sinh \Big(\frac{\lambda}{\lambda_z} (2k_y + q_y)
  z\Big)
  \int_0^{\infty} dz'\, e^{-(b_y/\lambda_z)\,z'}\, 2\sinh \Big(\frac{\lambda}{\lambda_z} (2k_y + q_y) z'\Big)
  \nonumber \\
&= \frac{1}{(2\pi)^2} \frac{q_x}{2\lambda q_x - \omega - i \delta} \, \frac{1}{24b_y^3 \lambda} \left[ b_y (b_y - \lambda q_y)
  \big( 2b_y^2 - 10 \lambda q_y b_y - 13 \lambda^2 q_y^2 \big) + 3\lambda^2 q_y^2 (8b_y^2 - \lambda^2 q_y^2)
    \arctanh \left(1-\frac{q_y \lambda}{b_y}\right) \right].
\end{align}
\end{widetext}
The other components of the response tensor are calculated from the spin-spin correlation and density-spin correlations in a similar fashion.

\section{Surface plasmons from FA states}
\label{app_plasmon}

Here, we present a quantum mechanical treatment of surface plasmon modes arising from the FA states \cite{Andolina2018}. We start from the observation that the total potential seen by a test charge in a WSM in response to an external potential $V_{\rm ext}(z,{\bf q},\omega)$ is given by
\begin{equation}
V_{\rm sc} (z,{\bf q},\omega) = V_{\rm ext} (z,{\bf q},\omega) + V_{\rm ind} (z,{\bf q},\omega),
\end{equation}
where ${\bf q} \equiv (q_x,q_y)$. The induced potential $V_{\rm ind} (z,{\bf q},\omega)$ is related to the induced carrier density as
\begin{equation}
V_{\rm ind} (z,{\bf q},\omega) = \int dz'\, v(z,z',{\bf q})\, n_{\rm ind}(z',{\bf q},\omega),
\end{equation}
where
\begin{equation}
v(z,z',{\bf q}) = \frac{2\pi e^2 }{\kappa q}\, \exp(-q|z-z'|)
\end{equation}
is the partial Fourier transform of the Coulomb potential. The induced carrier density, in turn, is related to the screened potential by
\begin{equation}
n_{\rm ind} (z,{\bf q},\omega) = - \int dz'\, \Pi_{00} (z,z',{\bf q},\omega)\, V_{\rm sc}(z',{\bf q},\omega).
\end{equation}
Therefore, we have 
\begin{align}
V_{\rm sc}(z,{\bf q},\omega) &= V_{\rm ext}(z,{\bf q},\omega) \nonumber \\
  {} - \int dz'& \int dz''\, v(z,z',{\bf q})\, \Pi_{00}(z',z'',{\bf q},\omega)\,
  V_{\rm sc}(z'',{\bf q},\omega) .
\end{align}
In the absence of an external potential, we have
\begin{align}
V_{\rm sc}(z,{\bf q},\omega) &= - \int dz' \int dz''\, v(z,z',{\bf q})\,
  \Pi_{00}(z',z'',{\bf q},\omega) \nonumber \\ 
&\quad{} \times V_{\rm sc}(z'',{\bf q},\omega),
\end{align} 
which is the condition for plasma oscillations. Plasmons are non-trivial solutions of this equation.

In the non-retarded regime of $c \gg \omega/q$, we use Poisson's equation to solve for the screened potential. In the absence of bulk carriers, we have $\nabla^2 V_{\rm sc} = (q^2 - \partial_z^2) V_{\rm sc} = 0$, assuming the FAs to contribute negligibly to the carrier density, i.e., for $|q_y| \ll b_y/2\lambda$ \footnote{Since $b_y/2\lambda$ is on the order of the inverse lattice constant this assumption is valid for a large range of $q_y$.}. Therefore, the screened potential can be written as
\begin{align}
V_{\rm sc} (z,{\bf q},\omega) = v_{\rm sc}\, ({\bf q},\omega) e^{-q|z|}.
\end{align}
Evidently, the electric field associated with the surface plasmon is localized at the surface with a decay length of~$1/q$.

\begin{figure}[tb]
\includegraphics[width=0.95\columnwidth]{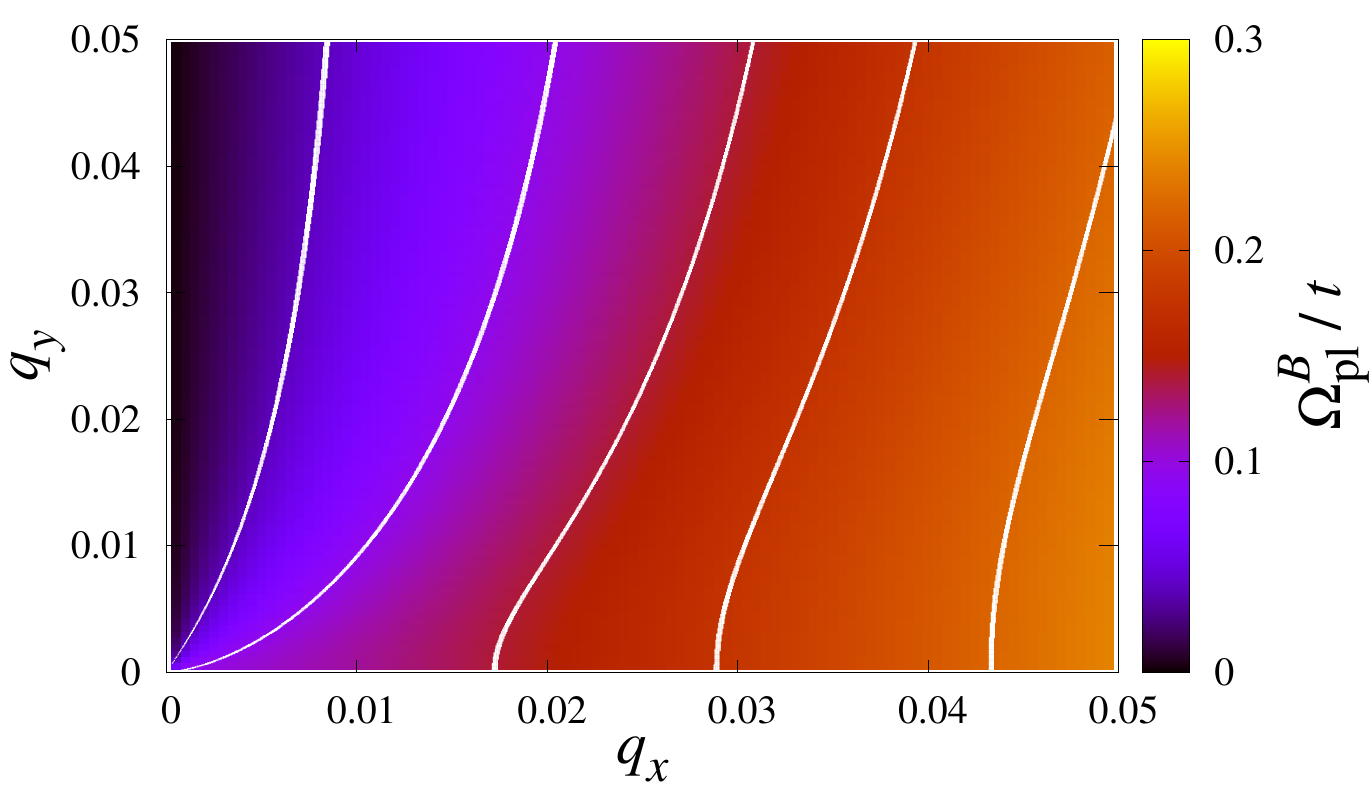}
\caption{Dispersion of the FA plasmons at the bottom surface, from Eq.\ (\ref{eq:plasmon_long}). The parameters are the same as in Fig.~\ref{fig_plasmon}.}
\label{fig_plasmon_extra}
\end{figure}

After some algebraic manipulations, the condition for the plasmon dispersion can be written as
\begin{align}
\epsilon_{\rm eff}^{T,B}\big({\bf q},\Omega_{\rm pl}^{T,B}\big) = 0,
\end{align}
where the effective surface dielectric constant reads
\begin{align}
\epsilon_{\rm eff}^{T,B}&({\bf q},\omega) = 1 + \frac{2\pi e^2 }{\kappa q} \nonumber \\
&{}\times \int dz' \int dz''\, \Pi_{00}^{T,B}(z',z'',{\bf q},\omega)\, e^{-q|z'+z''|},
\end{align}
and the real-space density response is described by
\begin{widetext}
\begin{align}
\Pi^{T,B}_{00}(z,z',{\bf q},\omega) &= - \frac{1}{(2\pi)^2} \sum_{\mu,\nu} \int dk_x \int dk_y \, \langle \phi^{\mu(T,B)} ({\bf k}_\parallel + {\bf q},z)|\sigma^0| \phi^{\mu(T,B)} ({\bf k}_\parallel,z) \rangle  \nonumber \\ 
 &\quad{}\times \langle \phi^{\nu(T,B)} ({\bf k}_\parallel,z')|\sigma^0| \phi^{\nu(T,B)} ({\bf k}_\parallel + {\bf q},z') \rangle \, \frac{n_F({\bf k}_\parallel)
    - n_F({\bf k}_\parallel + {\bf q})}
  {\epsilon({\bf k}_\parallel)
    - \epsilon({\bf k}_\parallel + {\bf q}) \pm (\omega + i \delta)} \nonumber \\ 
&= \frac{1}{(2\pi)^2} \frac{q_x}{2\lambda q_x \pm (\omega + i \delta)}\,
  \frac{1}{4b_y^2 \lambda_z^2} \int_{-{b_y}/{2\lambda}}^{{b_y}/{2\lambda} - q_y} dk_y\,
  (b_y^2 - 4 \lambda^2 k_y^2)\, \big[b_y^2 - 4 \lambda^2 (k_y + q_y)^2\big] \nonumber \\
&\quad{}\times e^{\pm (b_y/\lambda_z)\,z}\, 2\cosh \Big(\frac{\lambda}{\lambda_z}\, (2k_y + q_y)\,
  z\Big)\, e^{\pm(b_y/\lambda_z)\,z'}\, 2\cosh \Big(\frac{\lambda}{\lambda_z}\, (2k_y + q_y)\, z'\Big) ,
\end{align}
assuming $q_x,q_y > 0$.

After some tedious but straightforward calculations, the dispersion of FA plasmons is given by
\begin{align}
\Omega_{\rm pl}^{T,B} ({\bf q},\omega) &= \mp \cos\theta_{\bf q} \Bigg[ 2 \lambda q + \frac{\alpha_\kappa}{2\pi \lambda}\, \Bigg\{ b_y - 2\lambda |q_y| + \frac{\lambda^2 q_y^2}{2b_y}  + \frac{\lambda^3 |q_y|^3}{2b_y^2} + \frac{\lambda_z q}{2b_y^2}(b_y - \lambda |q_y|)(2b_y+ \lambda_z q)  \nonumber \\
&\quad{}+ \frac{ \lambda^4 q_y^4 - \lambda_z q (2b_y+ \lambda_z q) (4b_y^2 - 2\lambda^2 q_y^2 + 6 b_y \lambda_z q + 3\lambda_z^2 q^2) }{2b_y^2 (b_y+ \lambda_z q)}\, \arctanh\left(\frac{b_y - \lambda |q_y|}{b_y + \lambda_z q} \right) \Bigg\} \Bigg].
\label{eq:plasmon_long}
\end{align}
\end{widetext}
In the long-wavelength limit, $q \ll b_y/\lambda_z$, $|q_y| \ll b_y/2\lambda$, we obtain
\begin{align}
\Omega_{\rm pl}^{T,B} ({\bf q},\omega) = \mp \cos \theta_{\bf q} \left(
  \frac{\alpha_\kappa b_y}{\pi} + 2 \lambda q \right) ,
\end{align}
which is the same as obtained from the long-wavelength expansion of the linear density response.

The FA-plasmon dispersion calculated from Eq.\ (\ref{eq:plasmon_long}) is shown in Fig.\ \ref{fig_plasmon_extra}. The evident similarity to Fig.\ \ref{fig_plasmon} in terms of the plasmon dispersion as well as the constant-frequency contours confirms that the FA plasmons are well described by the long-wavelength expansion of the FA dielectric function described in Sec.~\ref{secVB1}.

\bibliography{Ghosh}

\end{document}